\documentclass[fleqn,10pt]{wlscirep}
\usepackage[utf8]{inputenc}
\usepackage[T1]{fontenc}

\usepackage{subcaption}

\usepackage{algorithmic}
\usepackage{algorithm}

\def\Underline{\setbox0\hbox\bgroup\let\\\endUnderline}
\def\endUnderline{\vphantom{y}\egroup\smash{\underline{\box0}}\\}
\def\|{\verb|}
\newcommand{\xx}{\mathbf{x}}

\newcommand{\uu}{\mathbf{u}}
\newcommand{\ww}{\mathbf{w}}
\newcommand{\vv}{\mathbf{v}}

\newcommand{\db}{\mathbf{d}}

\title{Application of QUBO solver using black-box optimization to structural design for resonance avoidance}

\author[1,*]{Tadayoshi Matsumori}
\author[1]{Masato Taki}
\author[1]{Tadashi Kadowaki}
\affil[1]{DENSO CORPORATION, 500-1, Minamiyama, Komenoki-cho, Nisshin, Aichi, 470-0111, Japan}

\affil[*]{tadayoshi.matsumori.j7b@jp.denso.com}

\begin{abstract}
Quadratic unconstrained binary optimization (QUBO) solvers can be applied to design an optimal structure to avoid resonance.
QUBO algorithms that work on a classical or quantum device have succeeded in some industrial applications.
However, their applications are still limited due to the difficulty of transforming from the original optimization problem to QUBO.
Recently, black-box optimization (BBO) methods have been proposed to tackle this issue using a machine learning technique and a Bayesian treatment for combinatorial optimization.
We propose a BBO method based on factorization machine to design a printed circuit board for resonance avoidance. 
This design problem is formulated to maximize natural frequency and simultaneously minimize the number of mounting points. 
The natural frequency, which is the bottleneck for the QUBO formulation, is approximated to a quadratic model in the BBO method. 
For the efficient approximation around the optimum solution, in the proposed method, we probabilistically generate the neighbors of the optimized solution of the current model and update the model. 
We demonstrated that the proposed method can find the optimum mounting point positions in shorter calculation time and higher success probability of finding the optimal solution than a conventional BBO method.
Our results can open up QUBO solvers' potential for other applications in structural designs.
\end{abstract}

\begin{document}
\flushbottom
\maketitle

\thispagestyle{empty}

\section{Introduction}

Computing algorithms and hardware that aim to solve a quadratic unconstrained binary optimization (QUBO) have been recently developed.
Quantum annealing (QA) \cite{kadowaki1998quantum, das2008colloquium} is one of the heuristic optimization algorithms for QUBO. 
QA utilizes quantum physics \cite{ray1989sherrington} to search for the optimal solution. 
QA is in the spotlight; after implementing QA in the quantum computer \cite{johnson2011quantum}, it was applied to some industry applications \cite{yarkoni2021quantum}.
In the last few years, heuristic optimization algorithms that work on a classical computer, e.g., FPGA and GPU, have also been proposed for QUBO \cite{tsukamoto2017accelerator,goto2019combinatorial,yoshimura2013spatial,inagaki2016coherent,amplify,irie2021}.
At present, these algorithms deal with a large number of design variables compared to QA that works on the quantum computer, and, in such a case, overcomes QA in terms of the time to solutions \cite{ohzeki2019control}.
In the present paper, we call for an algorithm to solve QUBO, a ``QUBO solver'' without distinguishing whether the algorithm performs on a classical or quantum device. 

QUBO solvers succeeded in solving some combinatorial optimization problems which are not directly formulated in a QUBO.
It is known that some combinatorial optimization problems can be transformed into QUBO \cite{lucas2014ising}.
In addition, other transformation techniques for QUBO formulation have also been proposed, such as encoding techniques to a binary variable \cite{chancellor2019domain,tamura2021performance} and converting methods from higher-order polynomial to a quadratic model \cite{dattani2019quadratization}.
And, furthermore, many applications have been reported as the success cases of the QUBO formulation, including vehicle path generation \cite{neukart2017traffic,ohzeki2019control}, traffic signal control\cite{inoue2021traffic}, portfolio optimization\cite{rosenberg2016solving}, quantum chemistry\cite{streif2019solving}, and machine learning\cite{amin2018quantum}.

When a QUBO solver is applied to a structural design in the industry, the QUBO formulation becomes a bottleneck.
Simulations of differential equations based on the finite element method are widely used to design products accounting for shapes, materials, geometric constraints, and so on. 
When the design problem is formulated as the optimization problem, the differential equations and their solutions are dealt with by constraint and objective functions, respectively.
For example, in the topology optimization which is a structural design method \cite{bendsoe1988generating,deaton2014survey}, design variables represent the structure shape, and a partial differential equation constrained optimization is formulated. 
The bottleneck of the QUBO solver in the design optimization is that the constraints and objectives cannot be explicitly expressed as the quadratic polynomial of binary design variables.
Design methods using black-box optimization (BBO) have been proposed to tackle the bottleneck \cite{kitai2020designing,wilson2021machine,izawa2021continuous,koshikawa2021benchmark}.
The BBO encodes the implicit functions of the binary variables into a binary quadratic model using the factorization machine (FM) \cite{rendle2010factorization} and the Bayesian optimization of combinatorial structures (BOCS) \cite{baptista2018bayesian}.
This methodology is extended to applications such as the metamaterial design \cite{kitai2020designing,wilson2021machine}, the chemical structure design \cite{hatakeyama2021tackling} and the vehicle design \cite{koshikawa2021combinatorial}.

In the present paper, we present a QUBO formulation relying on BBO for structural design optimization.
As an application, we deal with a printed circuit board (PCB) design in a vehicle for avoiding resonance.
A PCB used in the power control unit of a hybrid vehicle is a key device to control electric power.
The resonance of the PCB causes defects of the bonded electric parts and the connectors on the PCB.
Therefore, the PCB tightens screws through the mounting holes so that the natural frequency of the PCB becomes high to avoid low-frequency resonance.
One of the solutions to maximize the natural frequency is to fix the PCB with as many screws as possible.
However, the number of mounting holes is restricted because there are many electric parts on the PCB, and the number of screws affects fabrication cost in mass production.
Consequently, the PCB design requires maximizing natural frequency and minimizing the number of mounting holes.
We formulated this design problem as a multi-objective optimization problem, which is solved by two classical methods, the weighted sum method and the $\varepsilon$-constraint method \cite{deb2001}.

\begin{figure}[htbp]
	\centering
	\includegraphics[keepaspectratio, scale=0.8]{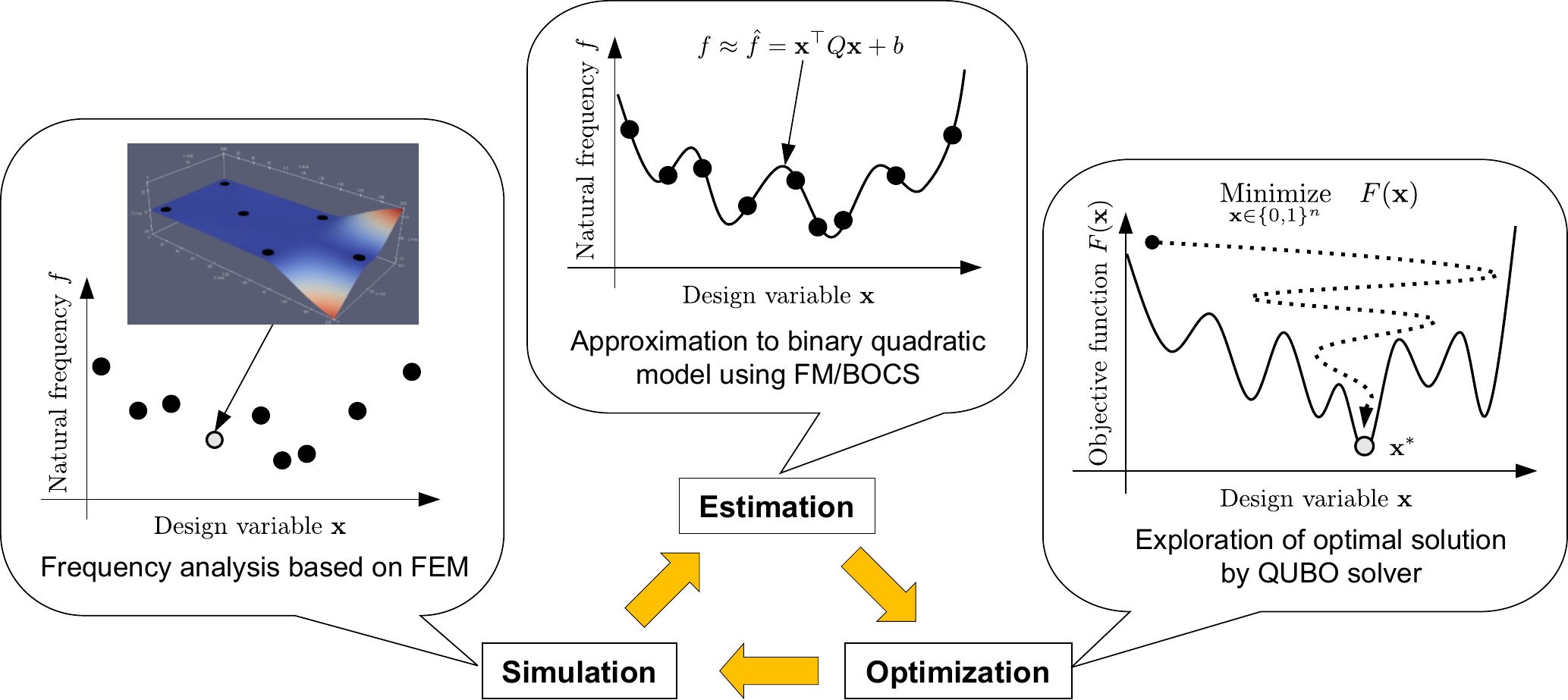}
	\caption{The schematic image of a black-box optimization for a QUBO solver for the printed circuit board design.}
	\label{fig:design_process}
\end{figure}

We demonstrated the performance of the FM-based black-box optimization for a QUBO solver in the PCB design.
Fig. \ref{fig:design_process} illustrates the PCB design procedure.
BBO approximates the natural frequency to a binary quadratic model followed by FMQA \cite{kitai2020designing}.
The natural frequency is a solution to the eigenvalue problem derived from the equation of motion.
We call the calculation of the eigenvalue problem a frequency analysis.
The eigenvalue problem and its solution were dealt with the constraint and objective in the optimization process.
Then we employed FM as BBO to approximate the natural frequency obtained from the frequency analysis to a binary quadratic model.
By doing this, the PCB design could be expressed in QUBO. 
In the present paper, the optimized solution to the QUBO and its neighbors which are probabilistically selected were utilized to update the BBO, while the original FMQA \cite{kitai2020designing} did not explicitly utilize a solution other than the optimized solution. 
The QUBO constructed by FM (hereinafter, our algorithm based on FMQA is called FM-QUBO) was applied to design PCB models with 17 or 27 candidates of the mounting holes. 
As a result, the proposed FM-QUBO showed higher success ratio of finding an optimal solution and shorter calculation time than the QUBO constructed by BOCS (BOCS-QUBO) \cite{koshikawa2021benchmark}. 
In addition, we confirmed that the selection of QUBO solvers, quantum annealing and simulated annealing, leaded to the difference of the success ratio.

\section{Results} \label{sec:results}

\subsection{Multi-objective optimization for resonance avoidance in PCB design}

The design of the mounting hole positions on the PCB to avoid resonance is formulated as a multi-objective optimization problem (MOOP) .
When the natural frequency $f$ of the PCB takes a high value, this design objective is satisfied.
In addition, the number of mounting holes $N$ has to be minimized to save space on the PCB and reduce the production cost.
Therefore, we formulate this design problem as a MOOP with two objectives.

The mounting hole positions were represented by binary design variables, $\xx=\{x_1,\ldots,x_n\}\in\{0,1\}^n$.
We prepared $n$ candidate positions of the mounting holes on the PCB design model.
Then, $x_i=1$ if position $i$ is selected as the mounting hole, $x_i=0$ if not. 
The number of the mounting holes $N$ can be calculated as the sum of $x_i$:
\begin{equation}
	N=\sum_{i=1}^n x_i. \label{eq:nom}
\end{equation}

The natural frequency $f$ depends on the mounting hole positions, i.e., the design variables $\xx$.
In the frequency analysis based on the finite element method, $f$ is calculated as a solution to an eigenvalue problem:
\begin{equation}
	G(\xx)=\left(K(\xx) - \lambda M(\xx) \right) \uu = 0 \label{eq:eigenfreq}
\end{equation}  
where $\lambda$ is the eigenvalue that is related to the natural frequency, i.e., $\lambda=\left(2\pi f\right)^2$, $\uu$ is the eigenvector, and $K$ and $M$ are the stiffness and mass matrix calculated from the finite element model of the PCB, respectively. 
In the present paper, we only dealt with the lowest natural frequency.
During the optimization, we assigned boundary conditions for fixing the displacement of a PCB model to the mounting hole positions corresponding to $\xx$.
Then, since $K$ and $M$ depend on where the boundary conditions are assigned, they have to be reassembled whenever $\xx$ changes.
Therefore, $G(\xx)$ is solved together with the optimization.

The MOOP for the PCB design can be formulated as
\begin{equation}
	\begin{aligned}
		\underset{\mathbf{x} \in \{0,1\}^n}{\text{Minimize}} && \{N,-f\},& \\
		\text{subject to} && G(\xx)=0.&
	\end{aligned} \label{eq:moop_original}
\end{equation}
We regard the eigenvalue problem, $G$, defined in Eq. \eqref{eq:eigenfreq}, as a constraint in the optimization.
A QUBO solver cannot solve this design problem directly because $f$ and $G$ is an implicit function of $\xx$.
Then, for expressing $f$ as the binary quadratic model of $\xx$ explicitly, we employed a machine learning technique and a Bayesian treatment, the factorization machine (FM) \cite{rendle2010factorization} and the Bayesian optimization of combinatorial structure (BOCS) \cite{baptista2018bayesian}.
Both approaches estimated the coefficients and constants of the binary quadratic model, $Q \in \mathbb{R}^{n\times n}, b\in \mathbb{R}$, based on data sets of $\xx$ and $f$, which are the inputs and outputs of the eigenvalue problem $G$.
The approximated natural frequency $\hat{f}$ is defined as
\begin{equation}
	f \approx \hat{f} = \xx^\top Q \xx + b. \label{eq:quad_form}
\end{equation}
For further details of the approximation calculation, see Method section \ref{subsec:BBO}.
Using $\hat{f}$, the optimization problem in Eq. \eqref{eq:moop_original} can be expressed by the quadratic polynomial of $\xx$ without the constraint: 
\begin{equation}
	\underset{\xx \in \{0,1\}^n}{\text{Minimize}} \quad \{N,-\hat{f}\}. \label{eq:moop_noconst}
\end{equation}

We have to convert the MOOP in Eq. \eqref{eq:moop_noconst} into QUBO, namely, two objectives into a single objective function $F(\mathbf{x})$:
\begin{equation}
	\underset{\xx \in \{0,1\}^n}{\text{Minimize}} \quad F(\mathbf{x}). \label{eq:qubo}
\end{equation}
The goal of the MOOP is to search for a set of optimal solutions, called Pareto-optimal solutions, which have trade-off relations.
We employed two classical multi-objective optimization methods: the weighted sum method and the $\varepsilon$-constraint method.
These methods scalarize multi-objectives into a single objective.

In the weighted sum method, the sum of the objectives with a weight $w\in[0,1]$ is utilized as the objective function:
\begin{equation}
	F(\mathbf{x}) = w\alpha N-(1-w)\beta\hat{f}, \label{eq:qubo_wsm}
\end{equation}
where $\alpha,\beta$ are parameters to normalize the objectives, $N$ and $\hat{f}$.
Equation \eqref{eq:qubo_wsm} has to be solved with different $w$ to obtain Pareto-optimal solutions.
This method can easily scalarize a set of objectives.
However, the drawback is that all optimal solutions cannot be found if the MOOP is non-convex \cite{deb2001}.

The $\varepsilon$-constraint method can be applied to the non-convex MOOP.
In this method, the objective functions except for one objective function deal with a constraint function, that is, 
\begin{equation}
	\begin{aligned}
		\underset{\mathbf{x} \in \{0,1\}^n}{\text{Minimize}} && -\hat{f},& \\
		\text{subject to} && N &= \overline{N}.
	\end{aligned}\label{eq:moop_ecm}
\end{equation}
When the MOOP in Eq. \eqref{eq:moop_ecm} is solved with all possible values of the parameter $\overline{N} \in \mathbb{Z}$, all Pareto-optimal solutions are found.
The above problem is converted into a QUBO with a penalty parameter $p$:
\begin{equation}
		F(\mathbf{x}) = - p \hat{f} + (N-\overline{N})^2, \label{eq:qubo_ecm}
\end{equation}
where $p$ takes an appropriate value so that the second term becomes larger than the first term.
One can also select the natural frequency $f$ as the constraint and minimize the number of mounting holes.
However, we considered that the formulation in Eq. \eqref{eq:moop_ecm} would suit the PCB design problem because the constraint value $\overline{N}$ is an integer value, and all Pareto-optimal solutions are found in $n$ times optimization.

\subsection{Algorithm of FM-based black-box optimization for QUBO solver}

The algorithm of FM-QUBO in a PCB design proceeds as shown in Algorithm \ref{alg1}. 
The algorithm basically follows FMQA \cite{kitai2020designing}.
One prepares a PCB model for the frequency analysis with sufficient initial and boundary conditions and assigns the initial values of the QUBO.
$n$ mounting hole candidate positions corresponding to the design variable $\xx$ are created on the PCB model.
The $m$ initial data sets of $\xx_i$ and the corresponding natural frequency $f_i$, $\db=\{d_1,\ldots, d_m\}, (d_i=\{\xx_i,f_i\}~i=1,\ldots,m)$, are prepared in advance to calculate the PCB model.
Then, the binary quadratic model in Eq. \eqref{eq:quad_form} that relies on FM is constructed to approximate $\db$. 
In the following, the QUBO in Eq. \eqref{eq:qubo} is solved by a QUBO solver and the optimized solution $\xx^*$ can be obtained.
We, in the present paper, prepared three solvers, simulated annealing (SA), quantum annealing (QA), and random search (RS).
After that, the frequency analysis is performed with $\xx^*$ to calculate the natural frequency $f^*$, and  $\db$ is updated.
Moreover, we create two solutions $\xx^\dagger, \xx^{\dagger\dagger}$ based on $\xx^*$ and added $\db$ as described below.
This process is iterated until the data set size $m$ reaches $\overline{m}$. 
Finally, a solution in $\db$ which minimizes the augmented objective function $F$ (Eq. \eqref{eq:qubo_wsm} or Eq. \eqref{eq:qubo_ecm}) is regarded as the best solution for this design problem.
We call these sequential procedures FM-QUBO and similar procedures using BOCS instead of FM BOCS-QUBO, respectively.
Note that, in BOCS-QUBO, the neighbors were not used in the data adding process, i.e., $m\leftarrow m+1$, because BOCS is a probabilistic algorithm, which estimates the distribution of quadratic models of random variables and then samples a model.

\begin{figure}[!t]
	\begin{algorithm}[H]
		\caption{FM-QUBO algorithm for resonance avoidance in the PCB design}
		\label{alg1}
		\begin{algorithmic}[1]
			\REQUIRE PCB design model with the $n$ candidates positions of mounting holes for the frequency analysis. $m$ initial data sets $\db$ of $\xx$ and their corresponding natural frequencies $f$. The maximum data size $\overline{m}$ for terminating the optimization. The parameters of the QUBO, and FM. $t=0, F^\prime=+\infty$. 
			\WHILE{$m<\overline{m}$}
			\STATE Construct the binary quadratic model (Eq. \eqref{eq:quad_form}) relying on FM using $\db$.
			\STATE Find $\xx^{*}$ for minimizing $F(\xx)$ (Eqs. \eqref{eq:qubo_wsm} or \eqref{eq:qubo_ecm}) by a QUBO solver.
			\STATE Select neighbors $\xx^\dagger, \xx^{\dagger\dagger}$ of $\xx^*$.
			\STATE Perform the frequency analysis with $\xx^*, \xx^\dagger$, and $\xx^{\dagger\dagger}$ and obtain their corresponding natural frequencies. 
			\STATE Update the data sets $\db$, $m\leftarrow m+3$. \label{alg1:update}
			\STATE Update the best solution, $\xx^\prime$, and its objective, $F^\prime$, if the objective of $\xx^{*}$, $\xx^\dagger$ or $\xx^{\dagger\dagger}$ is smaller than the current $F^\prime$.
			\ENDWHILE
			\RETURN The best mounting hole positions corresponding to $\xx^\prime$.
		\end{algorithmic}
	\end{algorithm}
\end{figure}

In FM-QUBO, the neighbors of the optimized solution $\xx^\dagger, \xx^{\dagger\dagger}$ are added to the optimized solution.
The data set size $\left|\db\right|$ at the beginning of the optimization is small.
When the optimized solution is only added, an optimized solution strongly depends on the initial data sets.
Then, we created two neighbor solutions $\xx^\dagger, \xx^{\dagger\dagger}$ of the optimized solution $\xx^*$ and added three data sets to $\db$, $m \leftarrow m+3$ in Algorithm \ref{alg1}, every time $\db$ is updated.
It is expected that the neighbors of $\xx^*$ take similar or smaller objective function values.
We define the neighbors based on the Hamming distance.
One or two variables are selected from the optimized solution, and the selected variables are inverted, i.e., 0 to 1 or 1 to 0.
This algorithm contributes to the search focusing on the solution space near the optimized solution, and the search for a global optimum solution due to its randomness.

In the present paper, the optimization was performed until the data set size reached $\overline{m}=1800$.
We conducted the calculation 20 times with different initial values of $\xx$.
The average objective function values of the best solution and their 95\% confidence interval were evaluated.
Note that the best solution represents the solution that takes the smallest objective function values in the obtained solutions.
In addition, we focus on how many calculations can find the optimal solution within 20 times calculation.
Then, the probability of finding the optimal solution is defined as the success ratio.

\subsection{PCB design based on the weighted sum method} \label{subsec:wsm}

\begin{figure}[htbp]
	\centering
	\includegraphics[keepaspectratio, width=7cm]{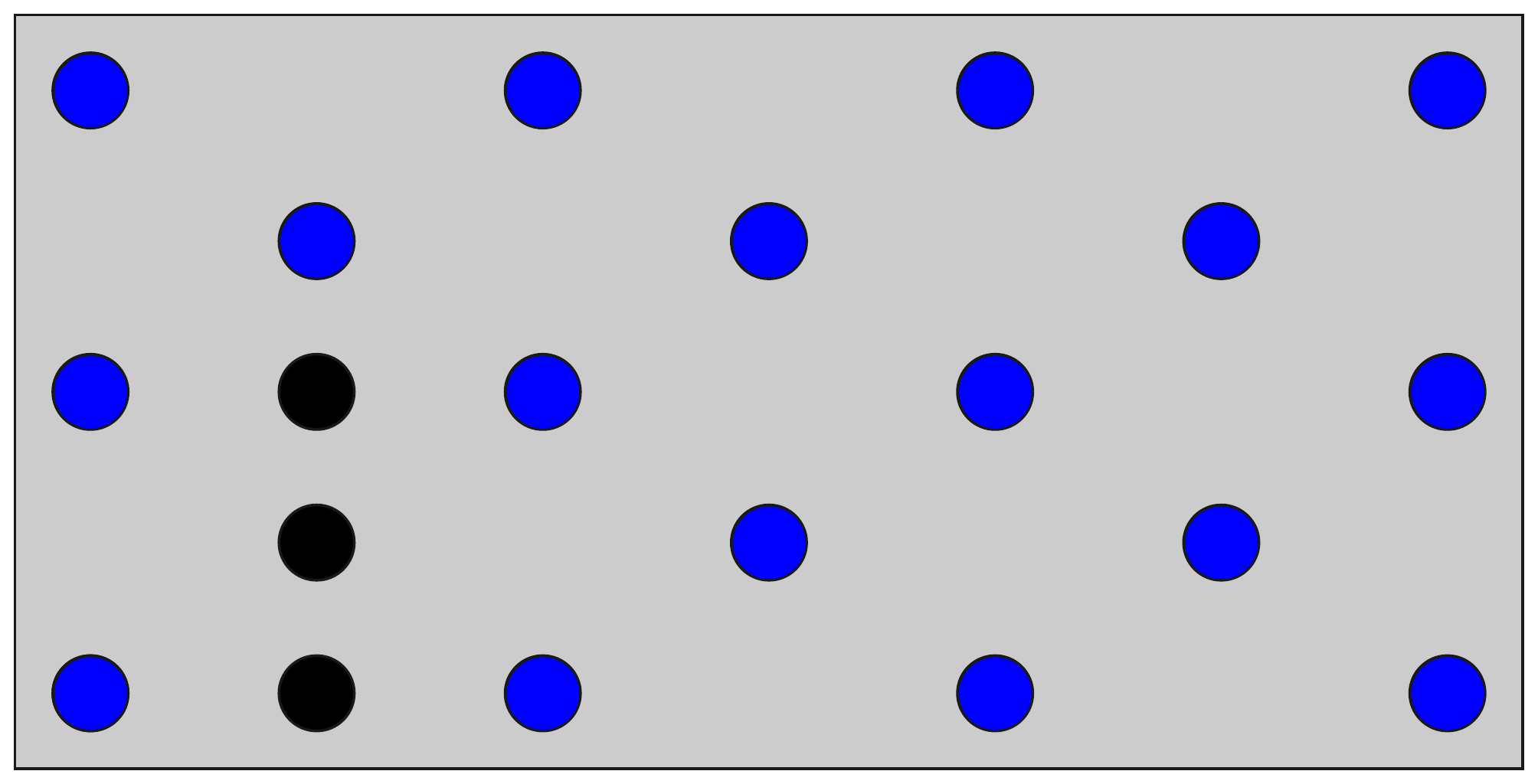}
	\caption{The simplified PCB model. The 17 blue circle areas show the candidates of the mounting holes. The additional masses are placed on the black circle areas.}
	\label{fig:17var_model}
\end{figure}

\begin{figure}[htbp]
	\centering
	\includegraphics[keepaspectratio, width=7cm]{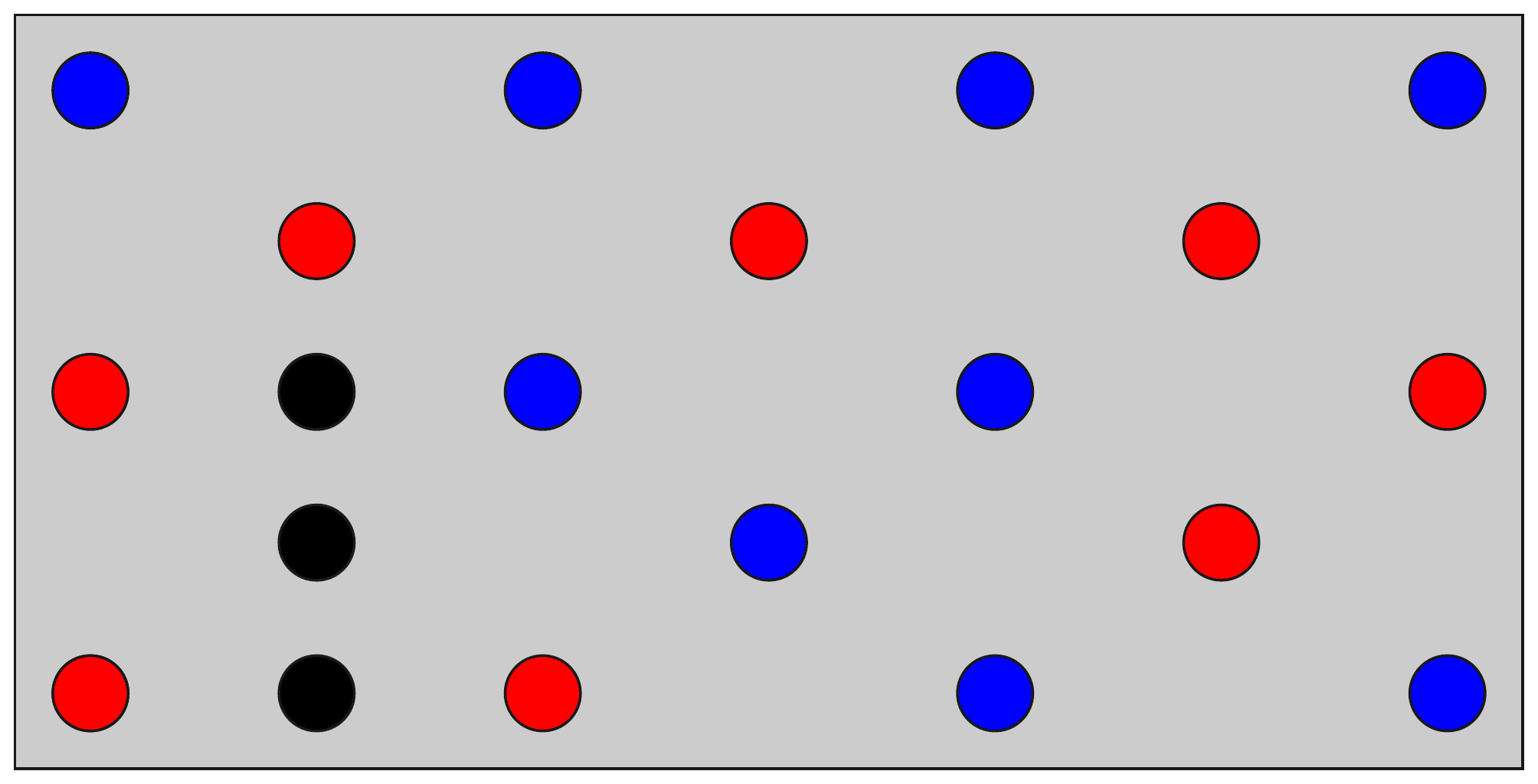}
	\caption{The optimal solution of the QUBO formulated by the weighted sum method (Eq. \eqref{eq:qubo_wsm}) with $w=0.5$. The red areas show the optimal mounting holes ($x_i=1$). The natural frequency $f$ and the number of mounting holes $N$ of this model are 825Hz and 8, respectively.}
	\label{fig:17var_wsm_opt}
\end{figure}

We prepared a simplified PCB model (Fig. \ref{fig:17var_model}) to demonstrate how the present algorithms work while designing the PCB model using the QUBO formulated by the weighted sum method as shown in Eq. \eqref{eq:qubo_wsm}.
This model has $n=17$ positions corresponding to mounting hole candidates, and three additional masses around the left corner.
The detail of the model parameters is shown in Method section \ref{subsec:freq}.
The number of combinations of the mounting holes is $2^{17}=131,072$.
The parameters $\alpha, \beta$ in Eq. \eqref{eq:qubo_wsm} were set to 1/450 and 1/8, respectively.
The parameter $w$ was set to $0.5$.
We computed the optimal solution of this problem setting using a brute-force search, as shown in Fig. \ref{fig:17var_wsm_opt}. 
The objectives of the optimal solution, i.e., the number of mounting holes $N^*$ and the natural frequency $f^*$, are $8$ and $825$Hz, respectively.
The optimal solution is one of the Pareto-optimal solutions.
Other Pareto-optimal solutions will be found when $w$ is changed in the range $[0,1]$.

\begin{figure}[htbp]
	\begin{minipage}[b]{0.45\linewidth}
		\centering
		\includegraphics[keepaspectratio, scale=0.55]{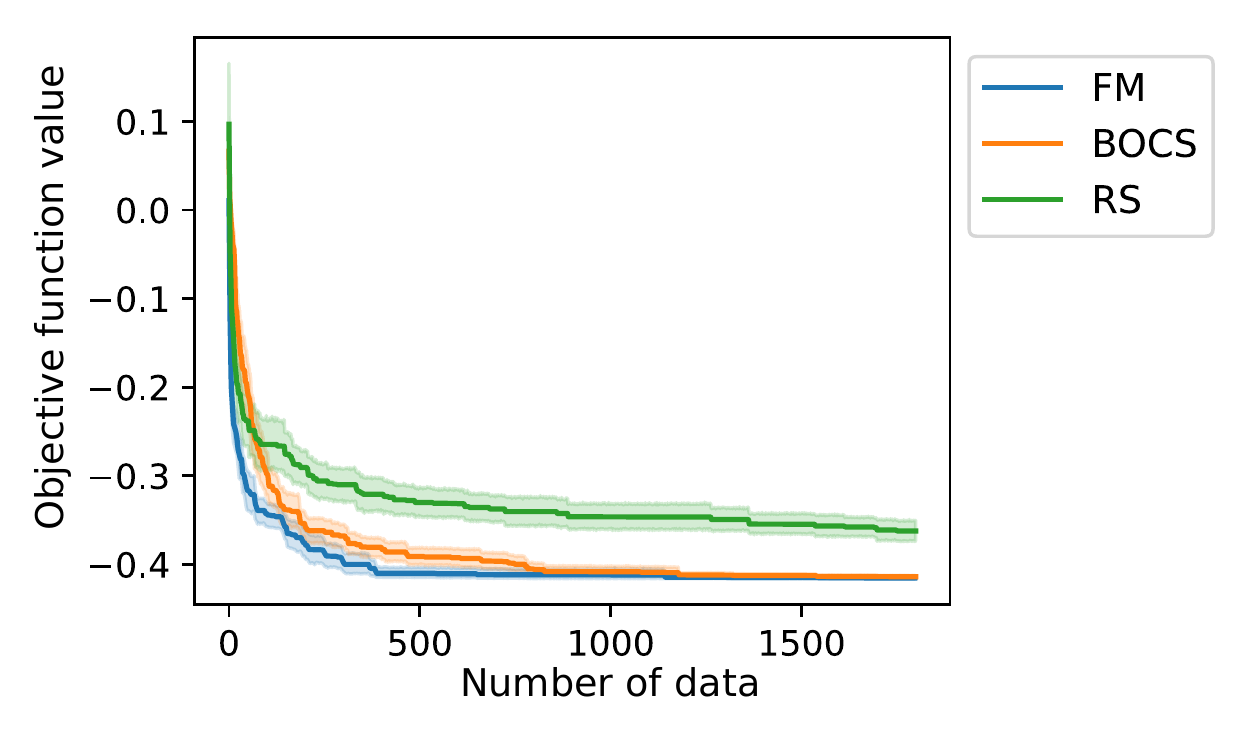}
		\subcaption{The best objective function value}
		\label{fig:17var_wsm_opt_hist_obj}
	\end{minipage}
	\begin{minipage}[b]{0.45\linewidth}
		\centering
		\includegraphics[keepaspectratio, scale=0.55]{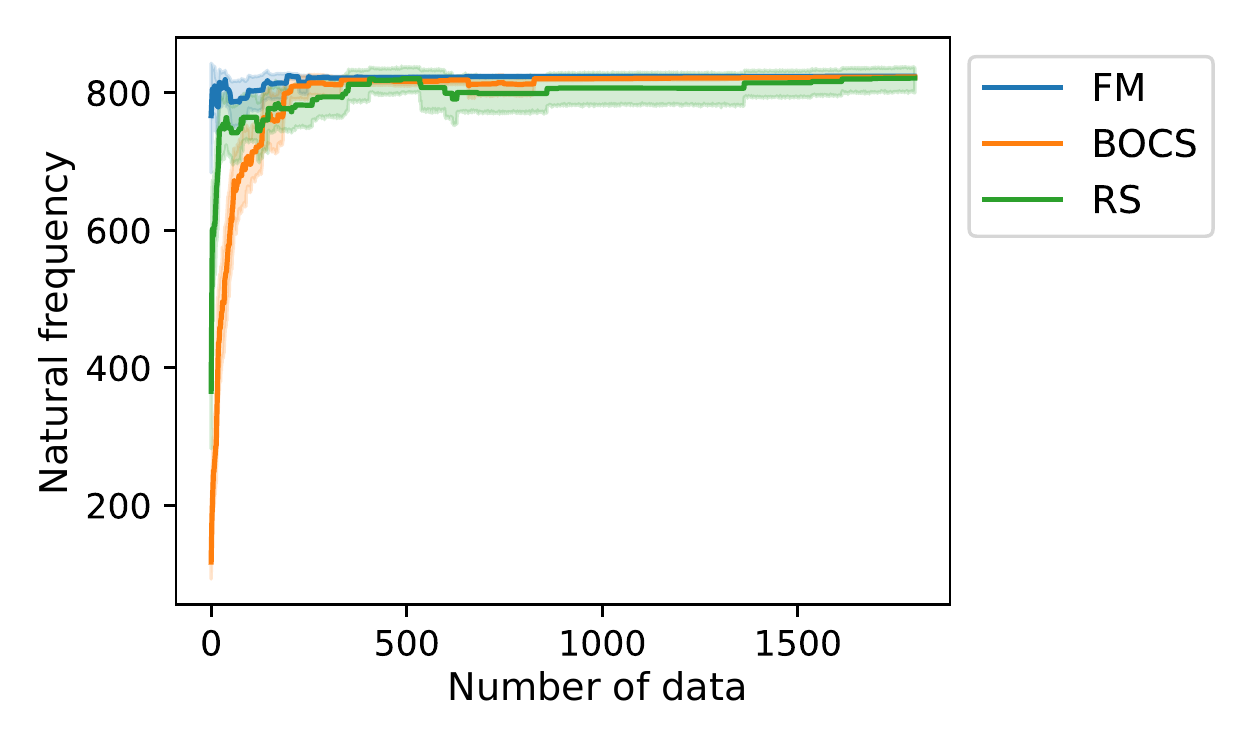}
		\subcaption{The best natural frequency}
		\label{fig:17var_wsm_opt_hist_freq}
	\end{minipage}\\
	\begin{minipage}[b]{0.45\linewidth}
		\centering
		\includegraphics[keepaspectratio, scale=0.55]{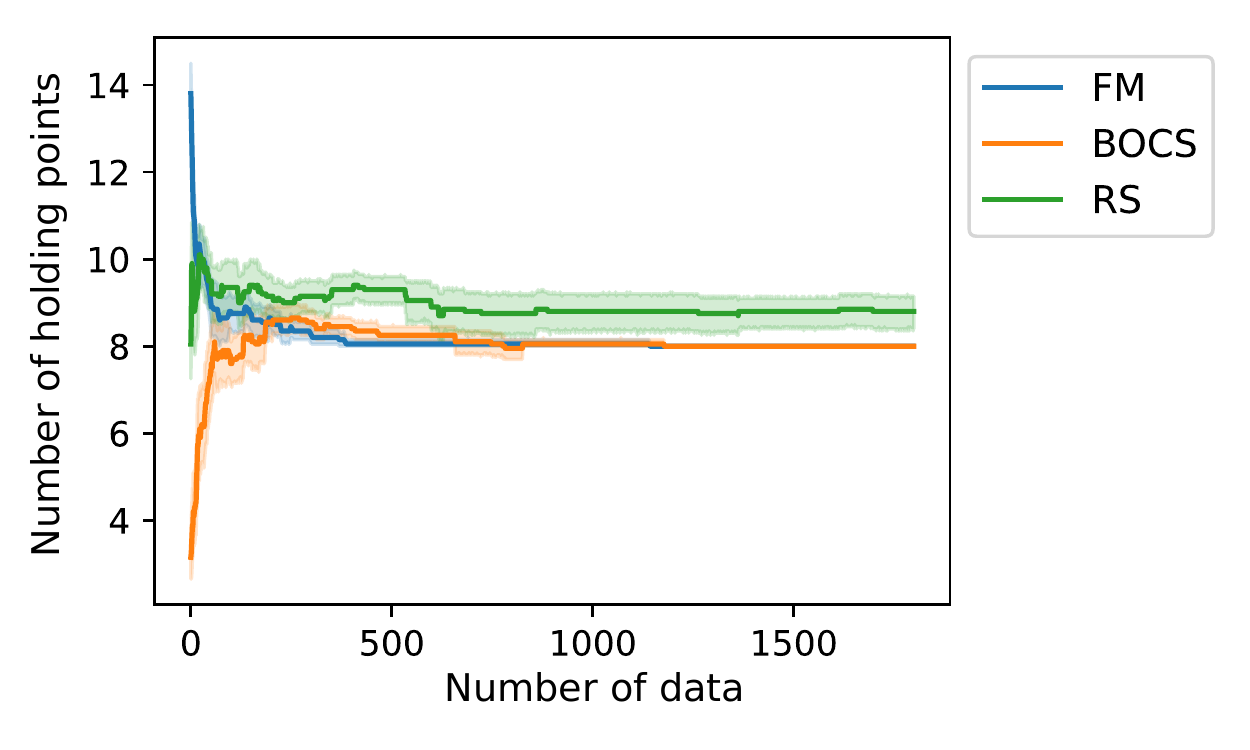}
		\subcaption{The best number of mounting holes}
		\label{fig:17var_wsm_opt_hist_pnts}
	\end{minipage}
	\caption{The optimization histories of the QUBO formulated by the weighted sum method (Eq. \eqref{eq:qubo_wsm}) with $w=0.5$ which was solved by FM-QUBO (FM) and BOCS-QUBO (BOCS). Random search (RS) was performed to compare their optimized results. The colored lines and areas illustrate the average objective function values of the best solution and their 95\% confidence interval when the QUBO was solved 20 times with different initial values.}\label{fig:17var_wsm_opt_hist}
\end{figure}

At the end of the optimization, the objective function values of the best solutions of the FM- and BOCS-QUBO were not identical as shown in Fig. \ref{fig:17var_wsm_opt_hist}, while FM-QUBO required a shorter calculation time than BOCS-QUBO.
The FM- and BOCS-QUBO explored a smaller objective function value than the random search.
In the early stage of the optimization process, i.e., the data size $m$ was small, the natural frequency and the number of the mounting holes showed a different trend in FM- and BOCS-QUBO (Figs. \ref{fig:17var_wsm_opt_hist_freq} and \ref{fig:17var_wsm_opt_hist_pnts}).
As for the objective function value (Fig. \ref{fig:17var_wsm_opt_hist_obj}), BOCS-QUBO was converged slowly compared with FM-QUBO. 
However, when $m$ became large, these values converged to a similar value regardless of the approximation method.
Regarding the calculation time except for preparing the binary quadratic model, BOCS-QUBO took about 1.7 times longer than FM-QUBO.

\begin{figure}[htbp]
	\centering
	\includegraphics[keepaspectratio, scale=0.55]{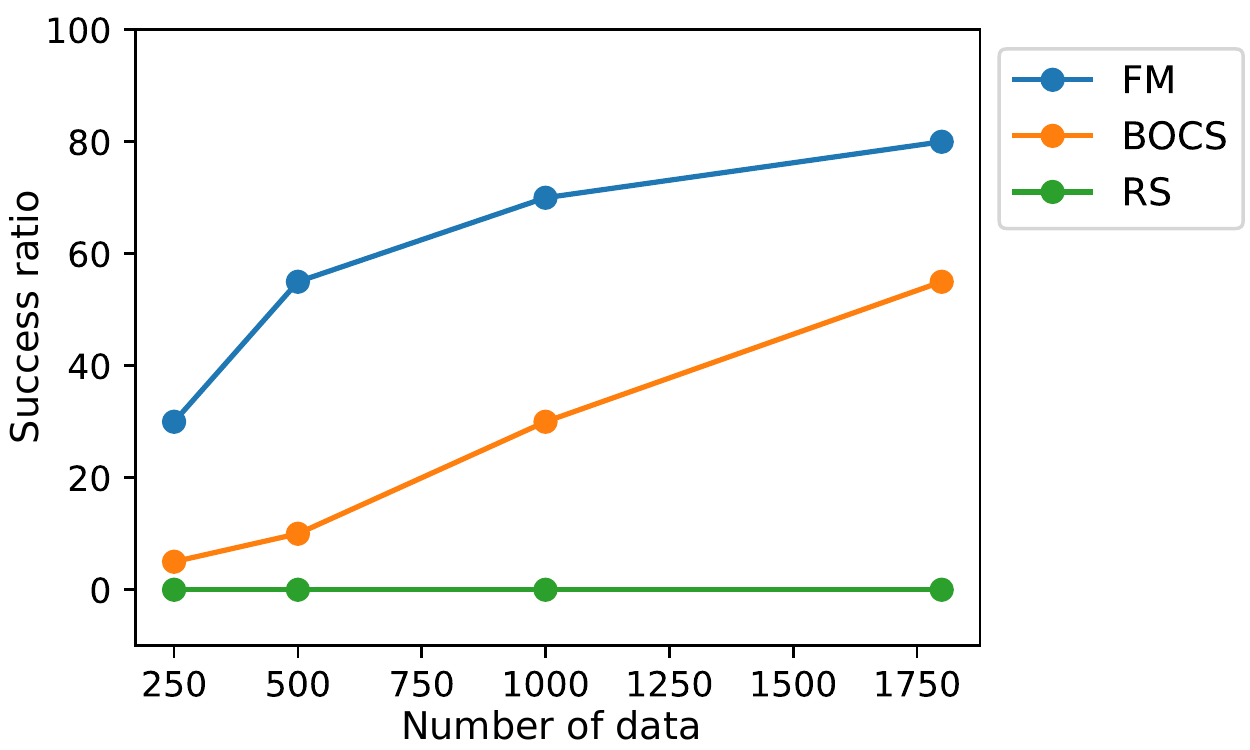}
	\caption{The success ratio of the QUBO formulated by the weighted sum method (Eq. \eqref{eq:qubo_wsm}). The success ratio is defined as the probability of the finding optimal solution when the QUBO was conducted 20 times with different initial values. The QUBO was solved by FM-QUBO (FM), BOCS-QUBO (BOCS) and the random search (RS). }
	\label{fig:17var_wsm_opt_num}
\end{figure}

FM-QUBO found the optimal solution using smaller data sets than BOCS-QUBO.
Fig. \ref{fig:17var_wsm_opt_num} shows the success ratio for different data sizes of $m=250, 500, 1000, 1800$, respectively.
When $m$ became large, FM- and BOCS-QUBO reached the optimal solution in high probability.
FM-QUBO obtained the optimal solution many times in the small $m$.
At the end of the optimization, i.e., the data size $m$ reached $\overline{m}=1800$, the success ratios were 80\% in FM-QUBO and 55\% in BOCS-QUBO.
The success ratio of BOCS-QUBO linearly increased with $m$.
Therefore, BOCS-QUBO may need more data sets to acquire the optimal solution in higher probability.

\subsection{PCB design based on the $\varepsilon$-constraint method} \label{subsec:ecm}

\begin{figure}[htbp]
	\centering
	\includegraphics[keepaspectratio, width=7cm]{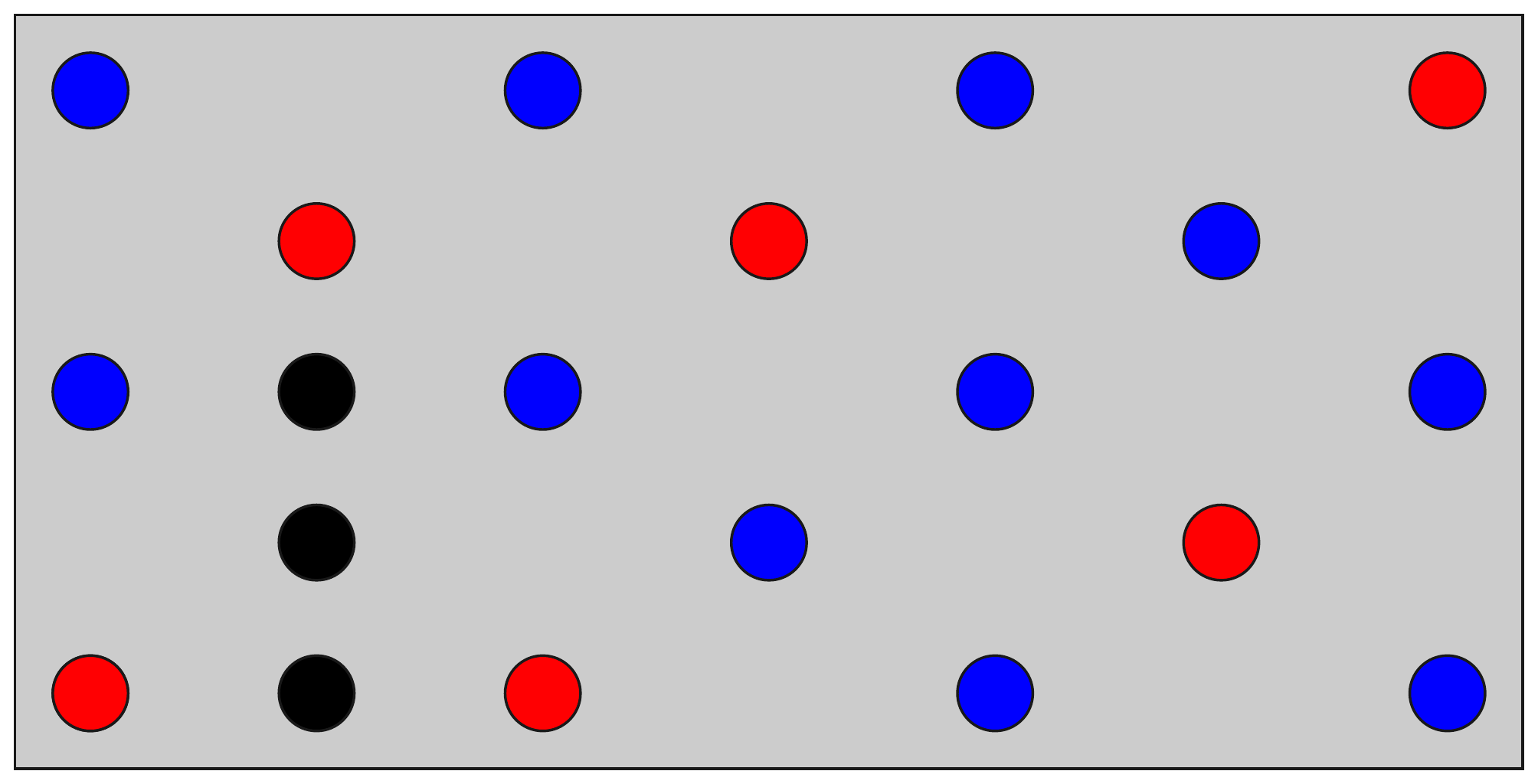}
	\caption{The optimal solution of the QUBO formulated by the $\varepsilon$-constraint method. The red areas show the optimal mounting holes. The natural frequency $f$ and the number of mounting holes $N$ of this model are 672Hz and 6, respectively. }
	\label{fig:17var_ecm_opt}
\end{figure}

The simplified PCB model (Fig. \ref{fig:17var_model}) was designed using the QUBO formulated by the $\varepsilon$-constraint method, as shown in Eq. \eqref{eq:qubo_ecm}.
We only show the result for the case of the constraint with the number of the mounting holes $\overline{N}$ of $6$.
The feasible solutions satisfying the constraint are ${}_{17}\mathrm{C}_6=12,376$.
The parameter $p$ in Eq. \eqref{eq:qubo_ecm} was set to 1/450.
The optimal solution is illustrated in Fig. \ref{fig:17var_ecm_opt}.

\begin{figure}[htbp]
	\begin{minipage}[b]{0.45\linewidth}
		\centering
		\includegraphics[keepaspectratio, scale=0.55]{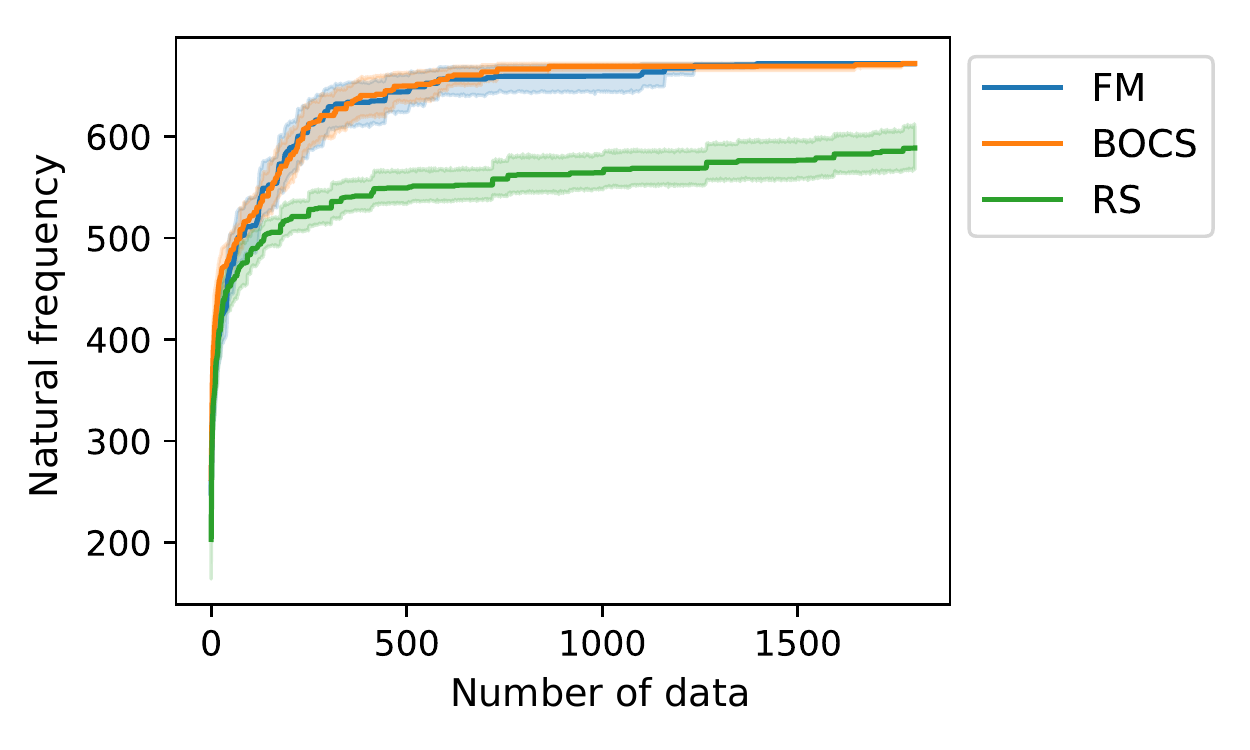}
		\subcaption{The natural frequency}
		\label{fig:17var_ecm_opt_hist_freq}
	\end{minipage}
	\begin{minipage}[b]{0.45\linewidth}
		\centering
		\includegraphics[keepaspectratio, scale=0.55]{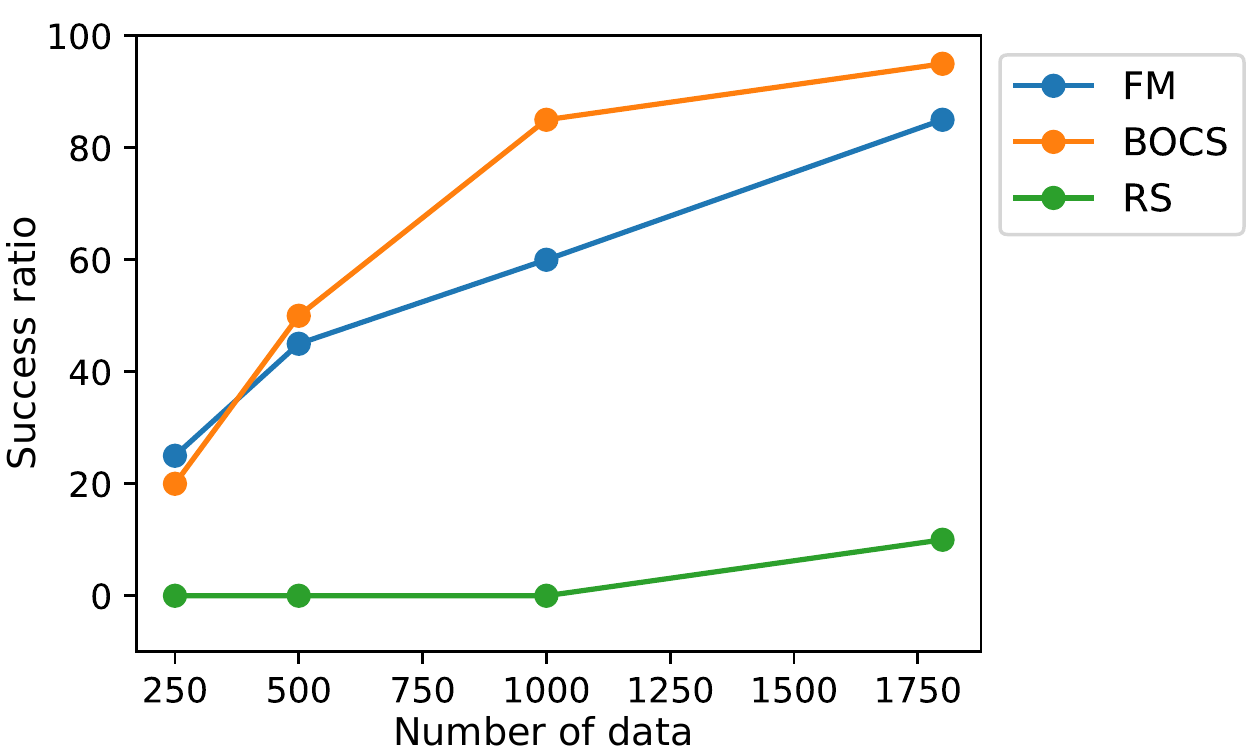}
		\subcaption{The success ratio}
		\label{fig:17var_ecm_opt_num}
	\end{minipage}
	\caption{(a) The optimization history of the QUBO formulated by the $\varepsilon$-constraint method (Eq. \eqref{eq:qubo_ecm}) with $\overline{N}=6$. (b) The success ratio at data size $m=250, 500, 1000, 1800$. The QUBO was conducted 20 times with different initial values using FM-QUBO (FM), BOCS-QUBO (BOCS) and random search (RS). }\label{fig:17var_ecm_opt_hist}
\end{figure}

During the optimization, the average natural frequencies of the optimized solution of FM- and BOCS-QUBO increased with the data size $m$ and demonstrated a similar trend, while that of random search gradually increased with $m$ (Fig. \ref{fig:17var_ecm_opt_hist_freq}).
At the end of the optimization, FM- and BOCS-QUBO could find the optimal solution of over 80\% (Fig. \ref{fig:17var_ecm_opt_num}).
In this problem setting, the success ratio of FM-QUBO was going up with $m$ after $m=1000$, but that of BOCS-QUBO was not the same. 
This result shows an opposite trend compared to the results given in Fig. \ref{fig:17var_wsm_opt_num}.

\begin{figure}[t!]
	\centering
	\begin{minipage}[b]{0.45\linewidth}
		\centering
		\includegraphics[keepaspectratio, width=7cm]{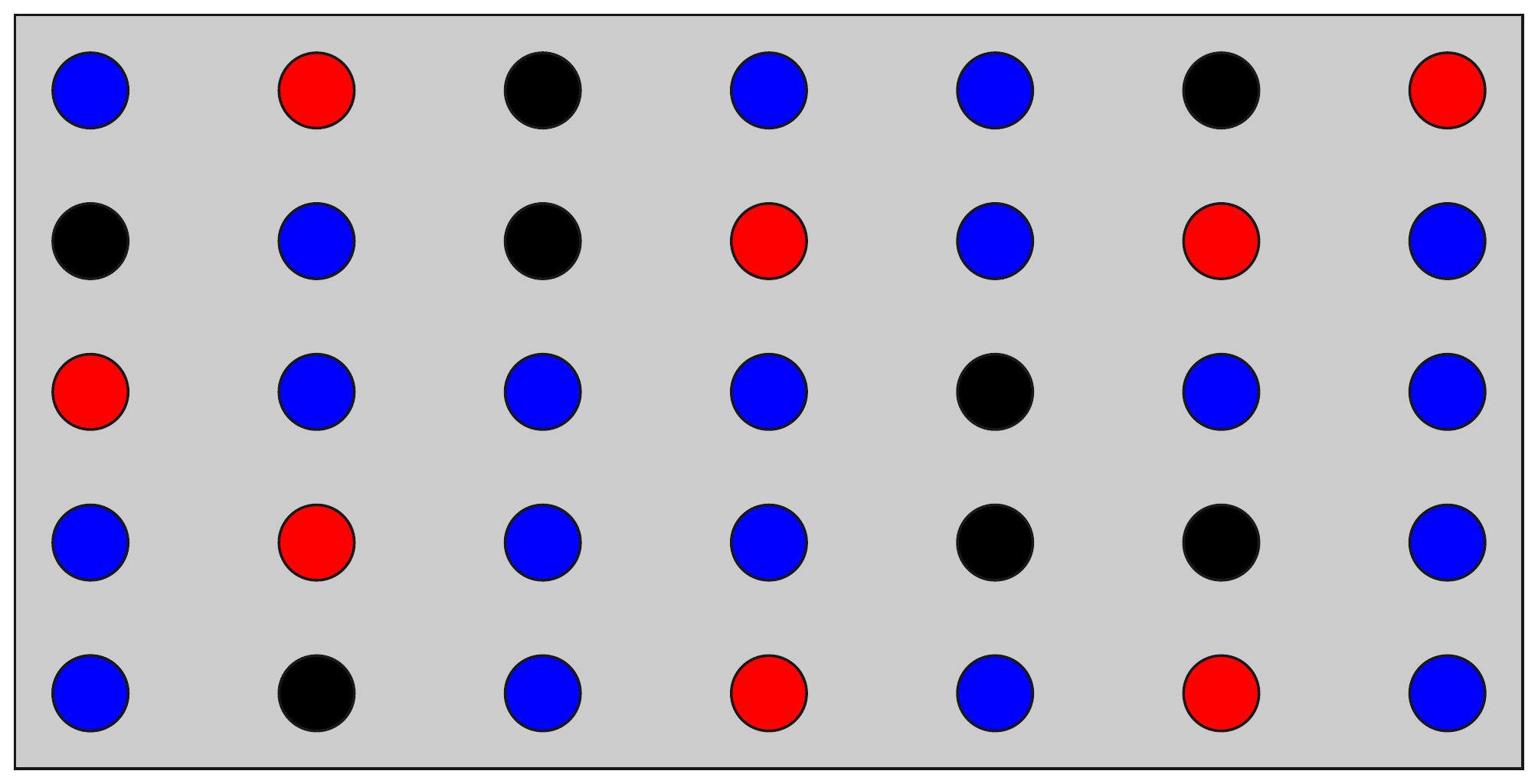}
		\subcaption{The optimized mounting holes}
		\label{fig:27var_model}
	\end{minipage}
	\begin{minipage}[b]{0.45\linewidth}
		\centering
		\includegraphics[keepaspectratio, scale=0.55]{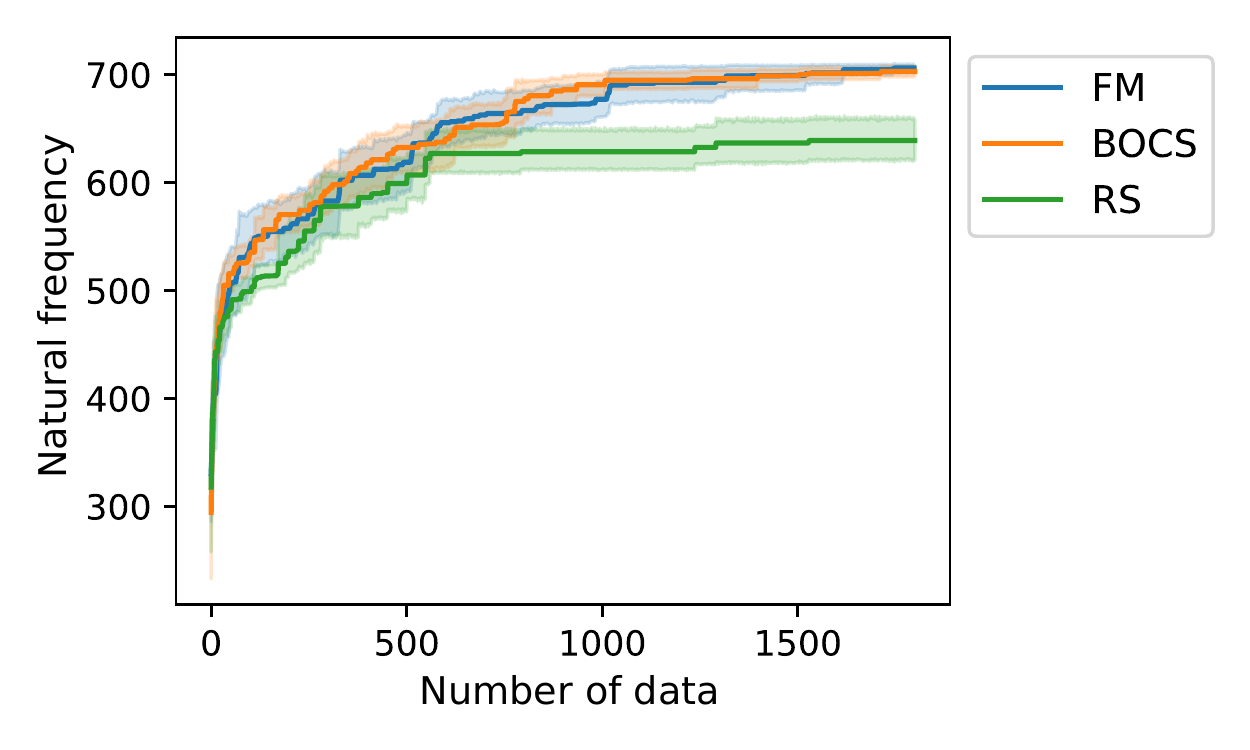}
		\subcaption{The best natural frequency}
		\label{fig:27var_ecm_opt_hist_freq}
	\end{minipage}
	\caption{The optimized result of the PCB model with 27 mounting hole candidates. The QUBO was formulated by the $\varepsilon$-constraint method (Eq. \eqref{eq:qubo_ecm}) with $\overline{N}=8$. (a) The optimized mounting holes (red circle areas) and the additional masses (black circle areas). Blue circles illustrate the candidates of the mounting holes which were not selected in optimization. (b) The average natural frequency of the best solution (colored solid lines) and their 95\% confidence interval (colored area) when the QUBO was conducted 20 times with different initial values using FM-QUBO (FM), BOCS-QUBO (BOCS) and random search (RS). }\label{fig:27var_ecm_opt_hist}
\end{figure}

We prepared another PCB model with 27 candidates of the mounting holes, i.e., $n=27$, and additional masses, as displayed in Fig. \ref{fig:27var_model}.
This design problem was also formulated as the QUBO based on the $\varepsilon$-constraint method with $\overline{N}=8$.
The number of combinations of the mounting holes is $2^{27}=134,217,728$, and the feasible solutions are ${}_{27}\mathrm{C}_8=2,220,075$.
In this large-scale problem, the difference in the best natural frequency between FM- and BOCS-QUBO could not be identified from the optimization history (Fig. \ref{fig:27var_ecm_opt_hist_freq}) as well as the result in the small-scale problem (Fig. \ref{fig:17var_ecm_opt_hist_freq}).
However, BOCS-QUBO took 35.9 times longer than FM-QUBO for preparing the binary quadratic model.

\section{Discussion}

We presented the FM-based black-box optimization for a QUBO solver to design the PCB for resonance avoidance.
The PCB design was formulated as the optimization problem with two objectives, the maximization of the lowest natural frequency and the minimization of the number of the mounting holes.
The multi-objective optimization problem was scalarized based on the classical multi-objective optimization techniques to formulate the PCB design as the QUBO: the weighted sum method and the $\varepsilon$-constraint method. 
In the scalarized optimization problem, a BBO method using factorization machine (FM) with the probabilistic data selection approximated the natural frequency which could not be directly expressed as the binary quadratic model.
The QUBOs constructed by FM (FM-QUBO) and a conventional BBO method using Bayesian optimization of combinatorial structures (BOCS-QUBO) were applied to design the simplified PCB models with 17 or 27 candidates of the mounting holes.

In our prepared design problems, FM-QUBO could find the optimal solutions in the high probability in the problems with both small and large solution spaces, and require shorter calculation time than BOCS-QUBO.
The average performance between FM- and BOCS-QUBO (Figs. \ref{fig:17var_wsm_opt_hist_obj}, \ref{fig:17var_ecm_opt_hist_freq}, and \ref{fig:27var_ecm_opt_hist_freq}) were not identical, while the success ratio (Figs. \ref{fig:17var_wsm_opt_num} and \ref{fig:17var_ecm_opt_num}) and the calculation time were different.
FM-QUBO explored the large solution space using the small data sets (Fig. \ref{fig:17var_wsm_opt_num}), and the elapsed time to estimate an approximation function was shorter than BOCS-QUBO.
In the case of a relatively small number of feasible solutions (Fig. \ref{fig:17var_ecm_opt_num}), FM- and BOCS-QUBO reached the optimal solution with a higher probability at the end of the optimization.
The difference in the calculation time directly depended on the time required for the parameter estimation in FM and BOCS.
The computational complexity in FM was less than $O(kn)$ where $k$ was a hyper-parameter of FM and was usually set to less than $n$ \cite{rendle2012factorization}, while in BOCS $O(n^3)$ \cite{baptista2018bayesian}.
Consequently, FM-QUBO overcomed BOCS-QUBO in terms of both the success ratio and calculation time.

\begin{figure}[htbp]
	\begin{minipage}[b]{0.9\linewidth}
		\centering
		\includegraphics[keepaspectratio, scale=0.55]{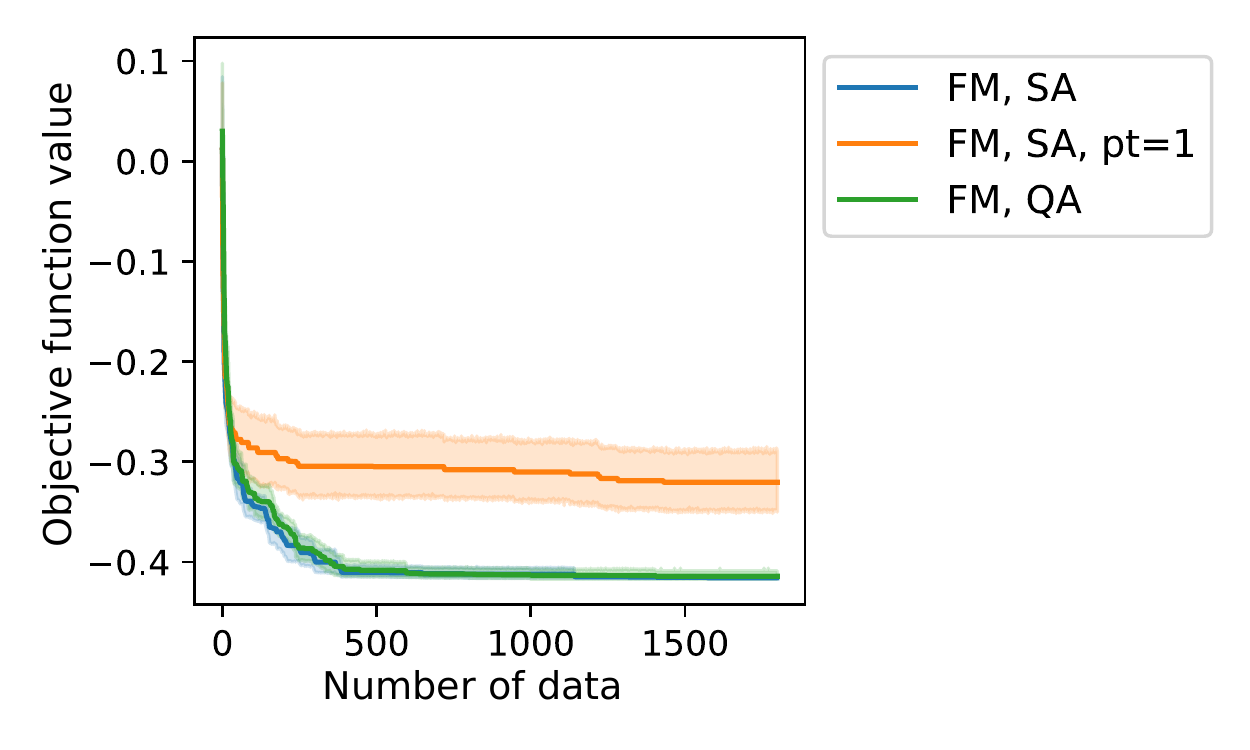}
		\includegraphics[keepaspectratio, scale=0.55]{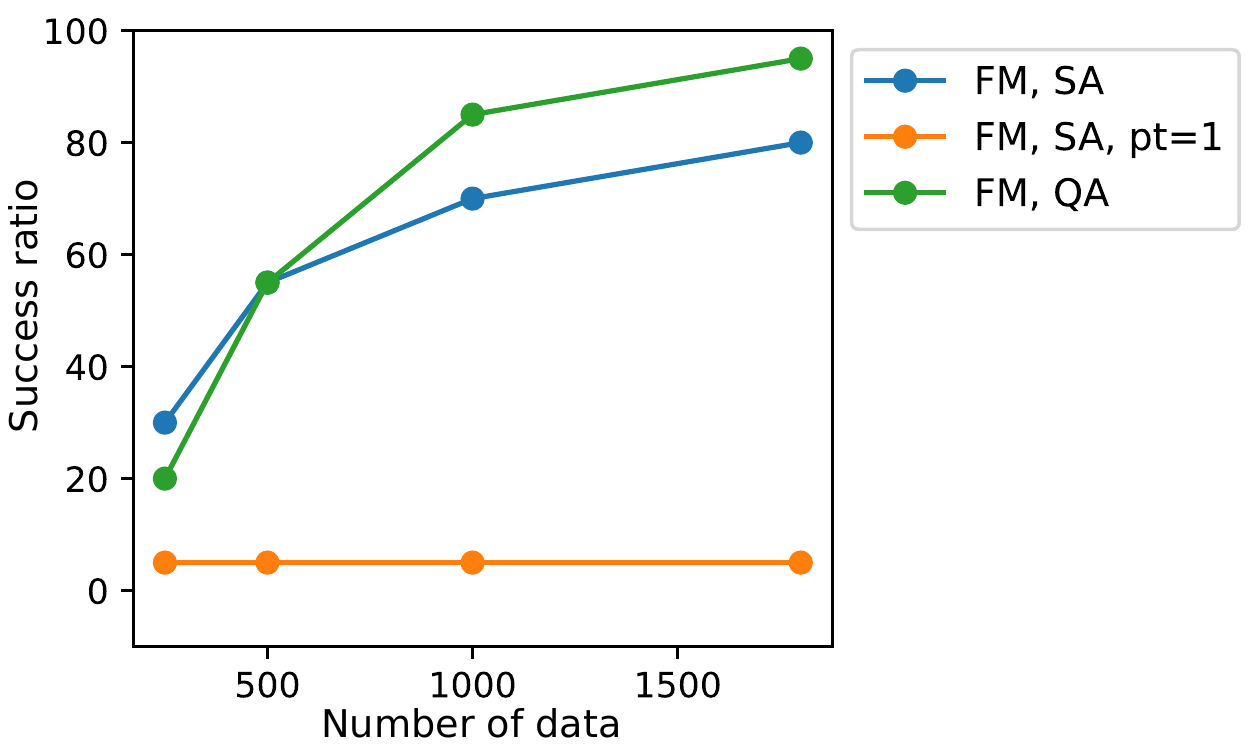}
		\subcaption{QUBO formulated by the weighted sum method (Eq. \eqref{eq:qubo_wsm})}
		\label{fig:17var_wsm_fm_qa}
	\end{minipage}
	\begin{minipage}[b]{0.90\linewidth}
		\centering
		\includegraphics[keepaspectratio, scale=0.55]{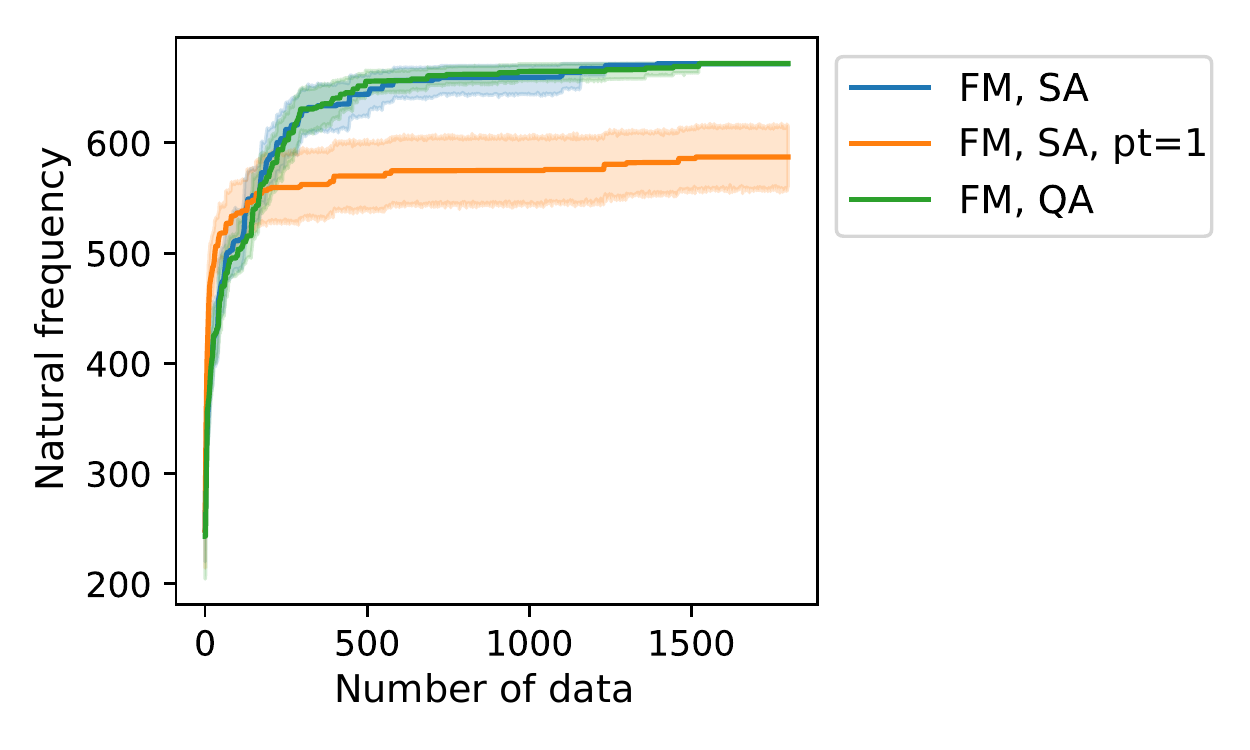}
		\includegraphics[keepaspectratio, scale=0.55]{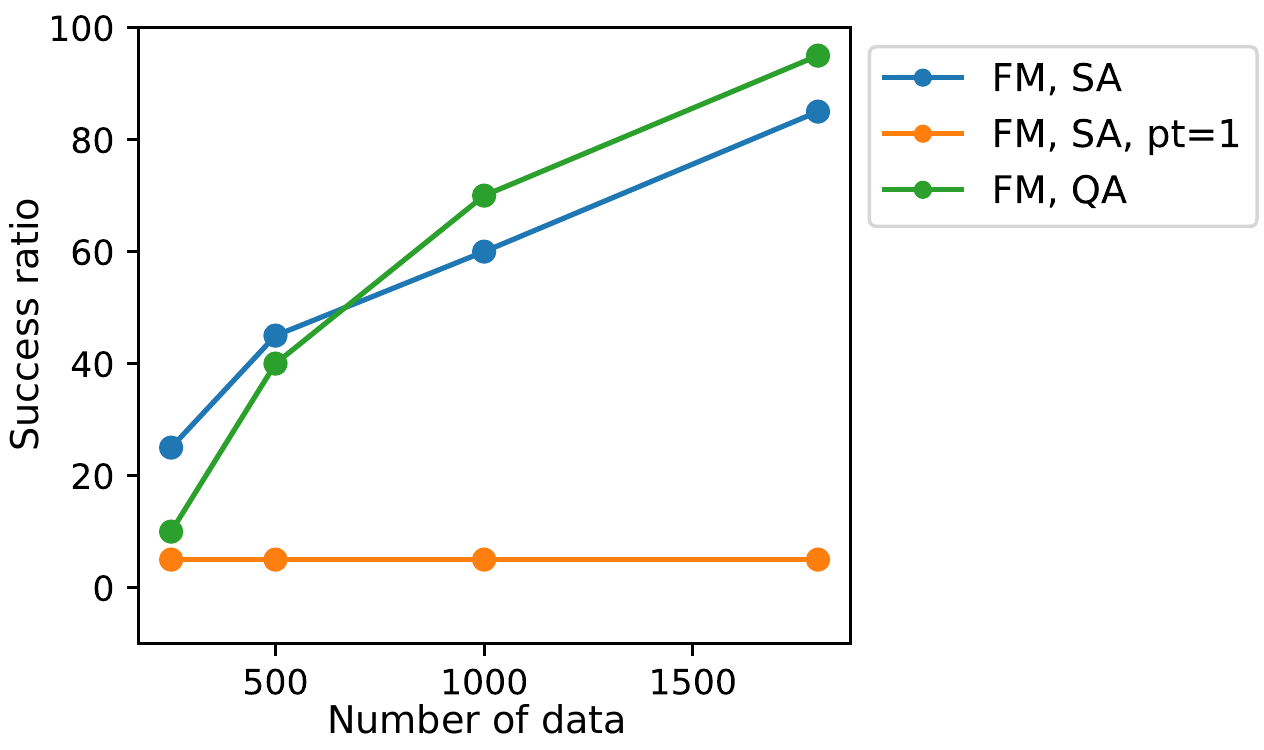}
		\subcaption{QUBO formulated by the $\varepsilon$-constraint method (Eq. \eqref{eq:qubo_ecm})}
		\label{fig:17var_ecm_fm_qa}
	\end{minipage}
	\caption{The optimized results of the FM-QUBO with 17 design variables using simulated annealing (SA) and quantum annealing (QA). We prepared two updating schemes. One was that three data sets which were the optimized result and its two neighbors (FM, SA) were added when FM was updated, the other was that one data set which was the optimized result (FM, SA, pt=1). The first column shows the average objective function values of the best solution and the second column the success ratio. The blue lines (FM, SA) are the same lines in Figs. \ref{fig:17var_wsm_opt_hist} and \ref{fig:17var_ecm_opt_hist}.}\label{fig:17var_fm_qa}
\end{figure}

To identify the effect of the proposed algorithm which is employed randomly selected neighbors of the optimized solution, we show the optimized results using a simple algorithm which only used the optimized solution when the data $\db$ was updated, i.e., $m \leftarrow m+1$, in Fig. \ref{fig:17var_fm_qa} (FM, SA, pt=1).
Under the same data size $\overline{m}=1800$, the simple algorithm only found the larger objectives' solution than the proposed algorithm and can reach the optimum solution at once within 20 times calculation. 
Therefore, we confirmed that the proposed algorithm with the probabilistic data selection contributed to finding a minimum solution and increasing the success ratio. 
Regarding the optimization problems with constraints, since the neighbors included the unfeasible solutions, in the worst case, one-third of the data sets for the approximation only satisfied the constraints, i.e., $N=\overline{N}$. 
Nevertheless, the proposed algorithm can find an optimum solution with higher success ratio comparable to BOCS-QUBO. 
We consider that the adding solutions violated the constraint condition would help precisely approximate the objective function $F$ including the constraint function in the QUBO formulation, and result in the efficient search for the optimum solution.

The selection of the QUBO solver will affect the optimized result.
The average performances of SA and QA were not identical, but QA explored the optimal solution at a slightly higher probability than SA (Fig. \ref{fig:17var_fm_qa}).
If this result only led to the QUBO solver's performance, the QA would be a suitable solver for black-box optimization for a QUBO solver (BBO-QUBO) with many design variables.
However, since we observed the small difference in the success ratio between SA and QA in a few examples, we cannot conclude that QA is a better solver than SA because heuristic optimization methods, including SA, utilize randomness for their efficient search in nature. 
In addition, we dealt with the small number of the design variables in the demonstrated problems. 
Therefore, in our problem setting, it may be difficult to find the advantage of QA regarding the calculation time as shown in the reference \cite{morita2006convergence}. 
When the design problem which has a large number of design variables is solved by the proposed method, the QA advantages will be discussed.

Finally, our results also suggest that the BBO-QUBO assists in expanding the QUBO solver's engineering applications.
The present paper employed the lowest natural frequency as the objective and was approximated to the quadratic model using BBO.
There are other design objectives in the PCB design, such as avoiding a specified frequency, reducing the deformation, and so on.
Then, the objectives are evaluated using lower eigenvalues and their eigenvectors.
As the eigenvalues are solutions to the same eigenvalue problem, the presented BBO-QUBO algorithms applies to these designs by modifying the objectives and constraints to be approximated.
In addition, by replacing the frequency anaysis with other simulations corresponding to the design problems, the BBO-QUBO would be widely used in engineering, although it is essential to confirm its performance in each problem.

\section{Methods}
\subsection{Approximation of natural frequency for QUBO formulation} \label{subsec:BBO}

\subsubsection{Factorization machine (FM)}

FM was developed for learning sparse data efficiently \cite{rendle2010factorization}.
Let introduce a binary quadratic model for the estimation of the observed data $y$:
\begin{equation}
	\hat{y} = w_0 + \sum_{i=1}^n w_i x_i + \sum_{i=1}^n \sum_{j=1}^n \sum_{l=1}^k v_{il} v_{jl} x_i x_j, \label{eq:fm}
\end{equation}
where $w_0 \in \mathbb{R}$ and $\ww=(w_1, \ldots, w_n )\in \mathbb{R}^n$ are the global bias and the strength of $i$-th variable.
$\vv_{i} = (v_{i1}, \ldots, v_{ik}) \in \mathbb{R}^{k}$ is the strength vector of interactions between $i$-th variable and the others.
The model parameters $w_0, \ww, \vv$ in Eq. \eqref{eq:fm} are estimated using a given data set of an input variable $\xx \in \{0,1\}^n$ and the corresponding output value $y \in \mathbb{R}$.
When there is enough data sets, the FM in Eq. \eqref{eq:fm} can approximate the interactions between the variables.
In general, $k$ is chosen sufficiently small because the number of the estimation parameters $\vv$ becomes small and the computation time can be reduced, $O(kn)$. 
However, our objective in using the FM is to construct a binary quadratic model from the data sets, namely, we are not supposed to set $k$ to a small value.

\begin{figure}[t!]
	\begin{minipage}[b]{0.90\linewidth}
		\centering
		\includegraphics[keepaspectratio, width=7cm]{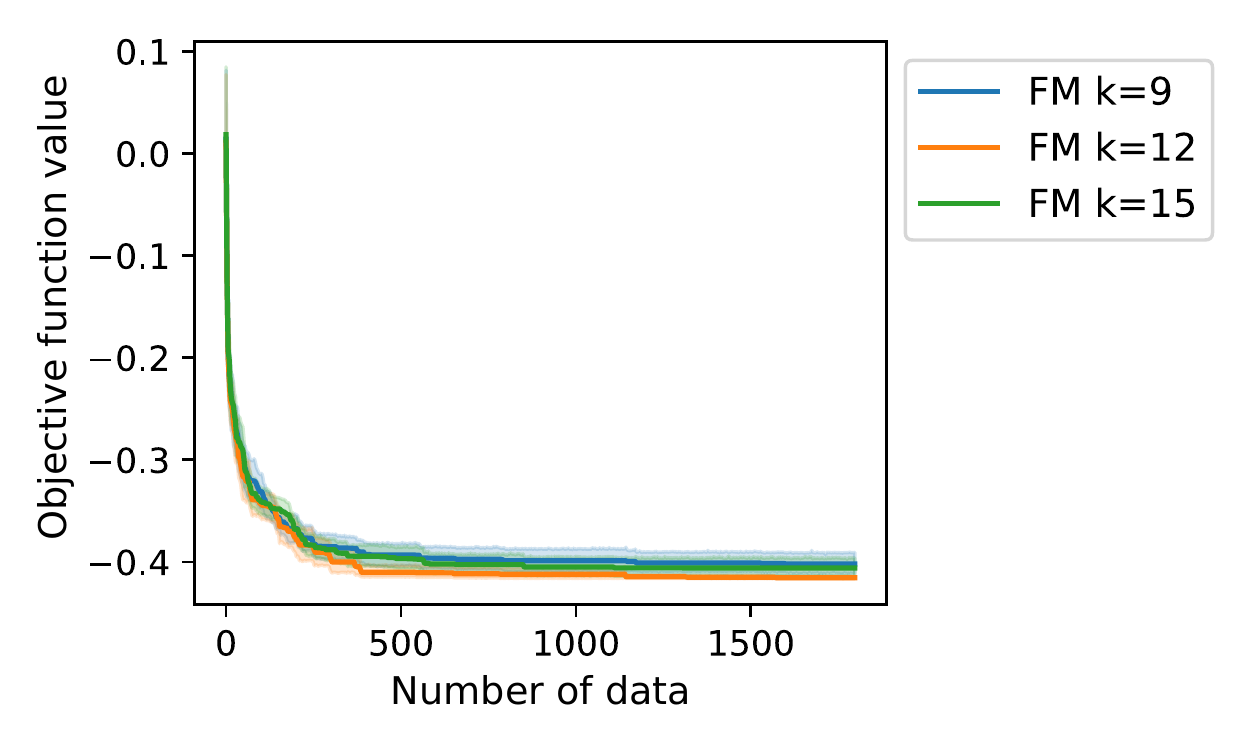}
		\includegraphics[keepaspectratio, width=7cm]{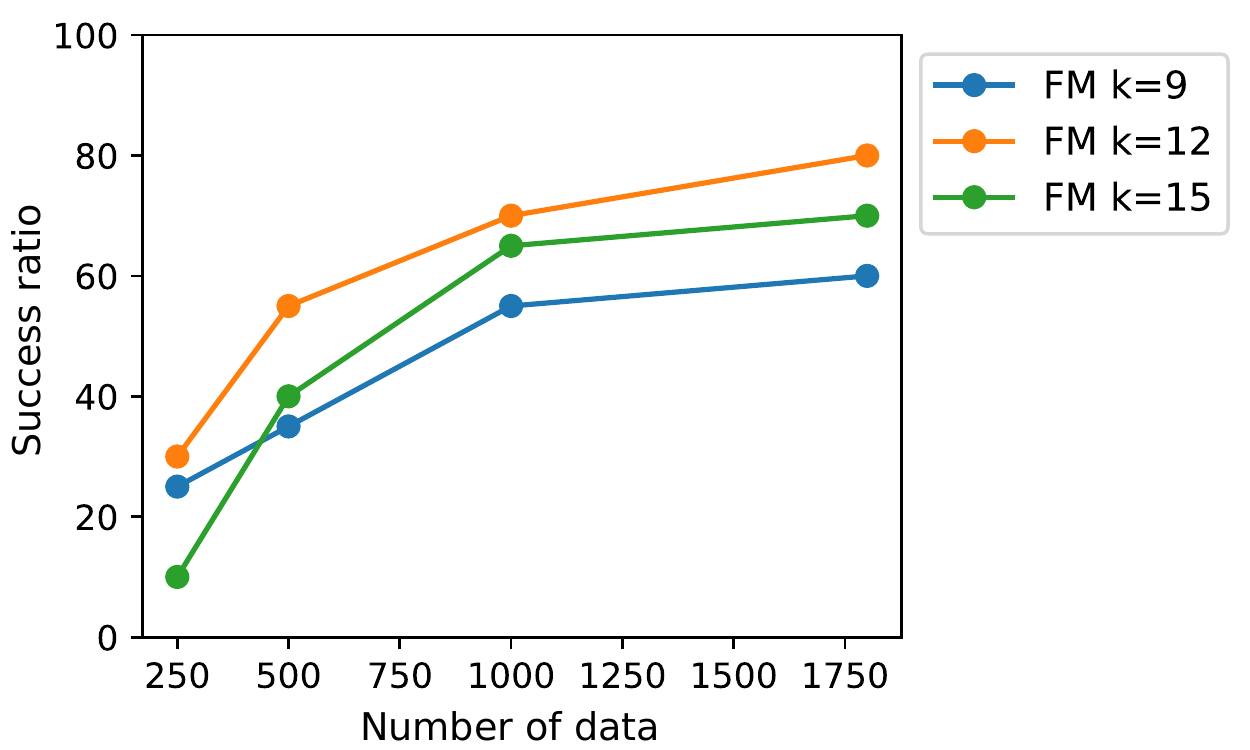}
		\subcaption{QUBO formulated by the weighted sum method}
		\label{fig:17var_wsm_fm_k}
	\end{minipage}\\
	\begin{minipage}[b]{0.90\linewidth}
		\centering
		\includegraphics[keepaspectratio, width=7cm]{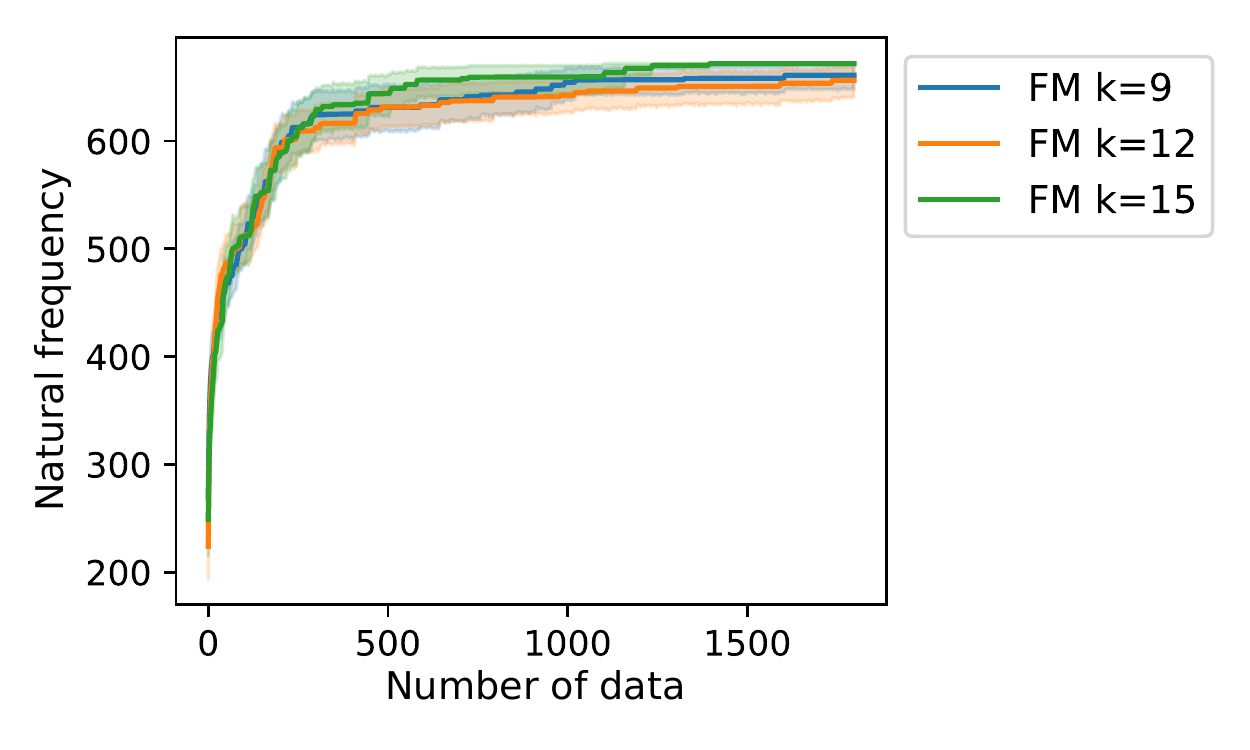}
		\includegraphics[keepaspectratio, width=7cm]{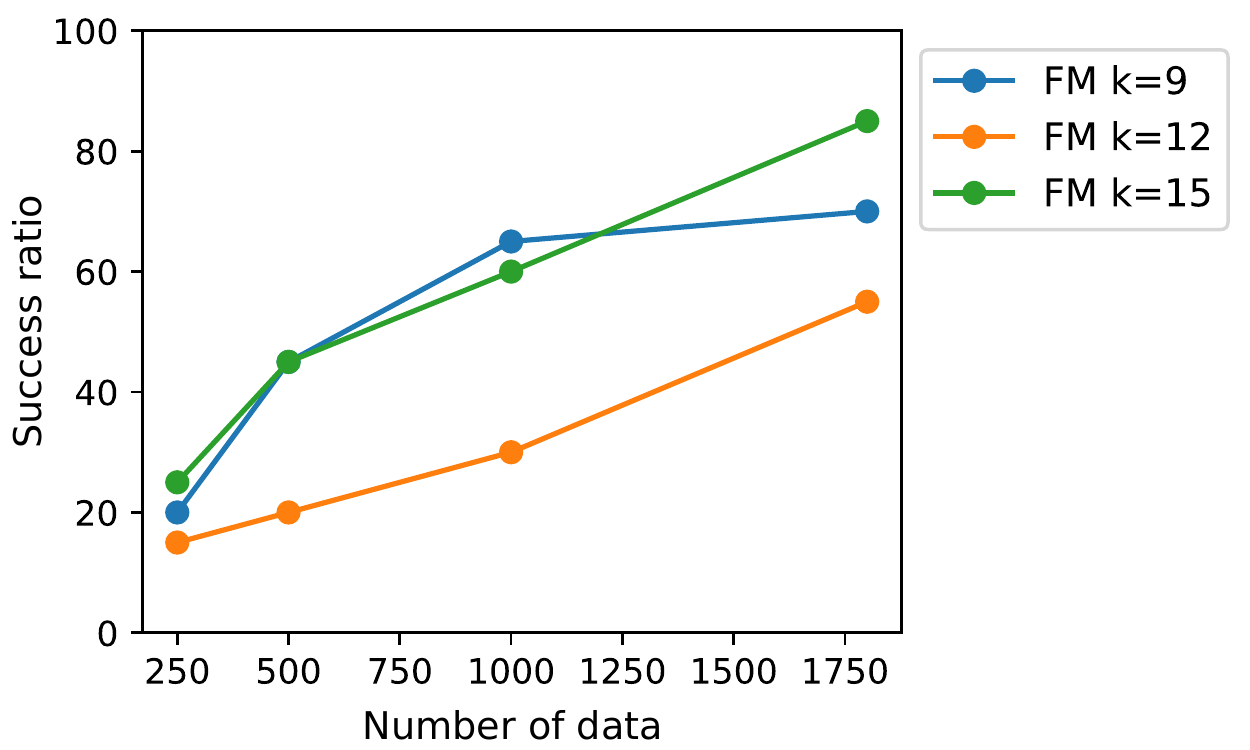}
		\subcaption{QUBO formulated by the $\varepsilon$-constraint method}
		\label{fig:17var_ecm_fm_k}
	\end{minipage}
	\caption{The optimized results of the QUBOs with $k=9,12,15$. The QUBOs were formulated by (a) the weighted sum method (Eq. \eqref{eq:qubo_wsm}) and (b) the $\varepsilon$-constraint method (Eq. \eqref{eq:qubo_ecm}) with 17 design variables. The results of $k=12$ in (a) and $k=15$ in (b) were same as the results in Figs. \ref{fig:17var_wsm_opt} and \ref{fig:17var_ecm_opt}. The first column shows the optimization history of the average objective function values of the best solution, and the second column shows the success ratio when the optimization was conducted 20 times with different initial values.}\label{fig:17var_fm_k}
\end{figure}

Before solving the QUBOs formulated by the weighted sum method (Eq. \eqref{eq:qubo_wsm}) and $\varepsilon$-constraint method (Eq. \eqref{eq:qubo_ecm}) as shown in Secs. \ref{subsec:wsm} and \ref{subsec:ecm}, we searched the proper value of $k$ in the QUBOs. 
We applied the FM-QUBO with different $k$ values, $k=9,12,15$, to the QUBOs and compared the average performances and the success ratios (Fig. \ref{fig:17var_fm_k}).
From the results, in our problem setting, we confirmed that $k$ affected the optimized results and the best $k$ value depended on the QUBO formulation.
Therefore, we employed the best $k$ in ``Results'', namely, $k=12$ in \ref{subsec:wsm} and $k=15$ in \ref{subsec:ecm}.
Note that the $k$ dependency was not observe in the original paper of FMQA \cite{kitai2020designing}.

\subsubsection{Bayesian optimization of combinatorial structures (BOCS)}

BOCS algorithm was proposed to solve the discrete optimization problem using scarce data.
An acquisition function $\hat{y}(\xx)$, which is an approximation function of observed data $y$ in BOCS, is defined as:
\begin{equation}
	\hat{y}(\xx)=\alpha_0 + \sum_j \alpha_j x_j + \sum_{i,j>i} \alpha_{ij} x_i x_j.
\end{equation}
The parameter $\mathbf{\alpha}=(\alpha_i,\alpha_{ij})$ is estimated by Bayesian treatment (see the reference \cite{baptista2018bayesian} for more details) using the data sets of $\xx$ and $y$.
As the Gibbs sampling is performed in the Bayesian treatment, the parameter estimation essentially consumes a long time, $O(n^3)$.
In general, a sufficiently large sampling size is required to approximate observed data precisely, and the size depends on the number of the design variables, $n$.

\subsection{Frequency analysis} \label{subsec:freq}

The frequency analysis was performed based on the finite element method to calculate the natural frequency of a PCB .
From the equation of motion of a PCB discretized by the finite element method, the eigenvalue problem can be derived as shown in Eq. \eqref{eq:eigenfreq}, i.e.,
\begin{equation*}
	\left( K(\mathbf{x}) -\lambda M(\mathbf{x})\right) \uu = 0.
\end{equation*}
$K \in \mathbb{R}^{n_\text{non}\times n_\text{non}}$ and $M \in \mathbb{R}^{n_\text{non}\times n_\text{non}}$ are the stiffness and mass matrix obtained when the equation of motion is discretized using finite elements. 
$n_\text{non}$ is the number of nodes of the discretized system.
$K$ and $M$ consists of the material properties, including density, Young's modulus, and Poisson ratio.
$\lambda$ and $\uu$ are the eigenvalue and the corresponding eigenvector.
$\lambda$ is related to the natural frequency $f$, i.e., $\lambda=\left(2\pi f\right)^2$.

\begin{figure}[htbp]
	\begin{minipage}[b]{0.45\linewidth}
		\centering
		\includegraphics[keepaspectratio, width=7cm]{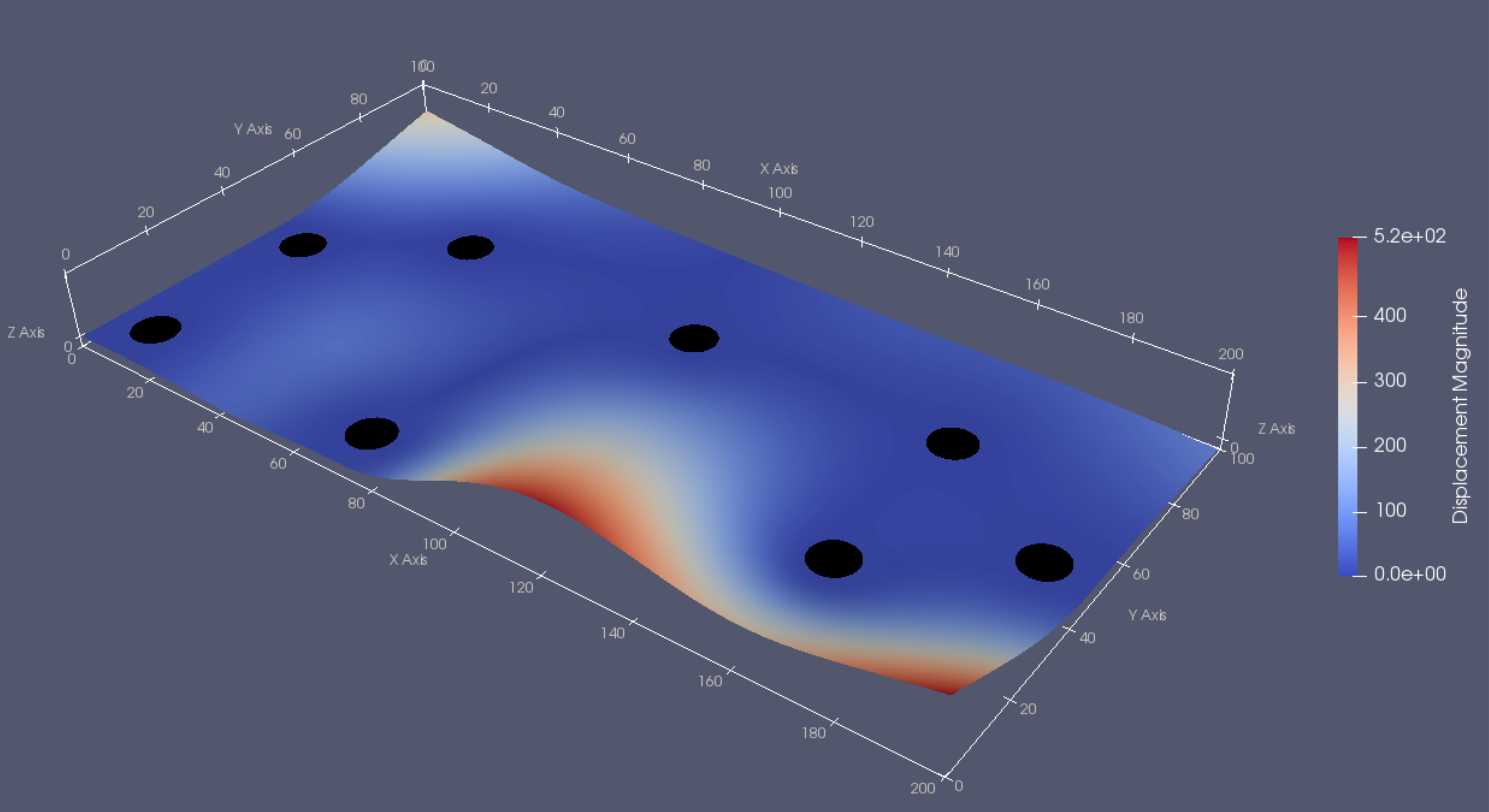}
		\subcaption{The QUBO in Eq. \eqref{eq:qubo_wsm} with $w=0.5$}
		\label{fig:17var_wsm_opt_mode}
	\end{minipage}
	\begin{minipage}[b]{0.45\linewidth}
		\centering
		\includegraphics[keepaspectratio, width=7cm]{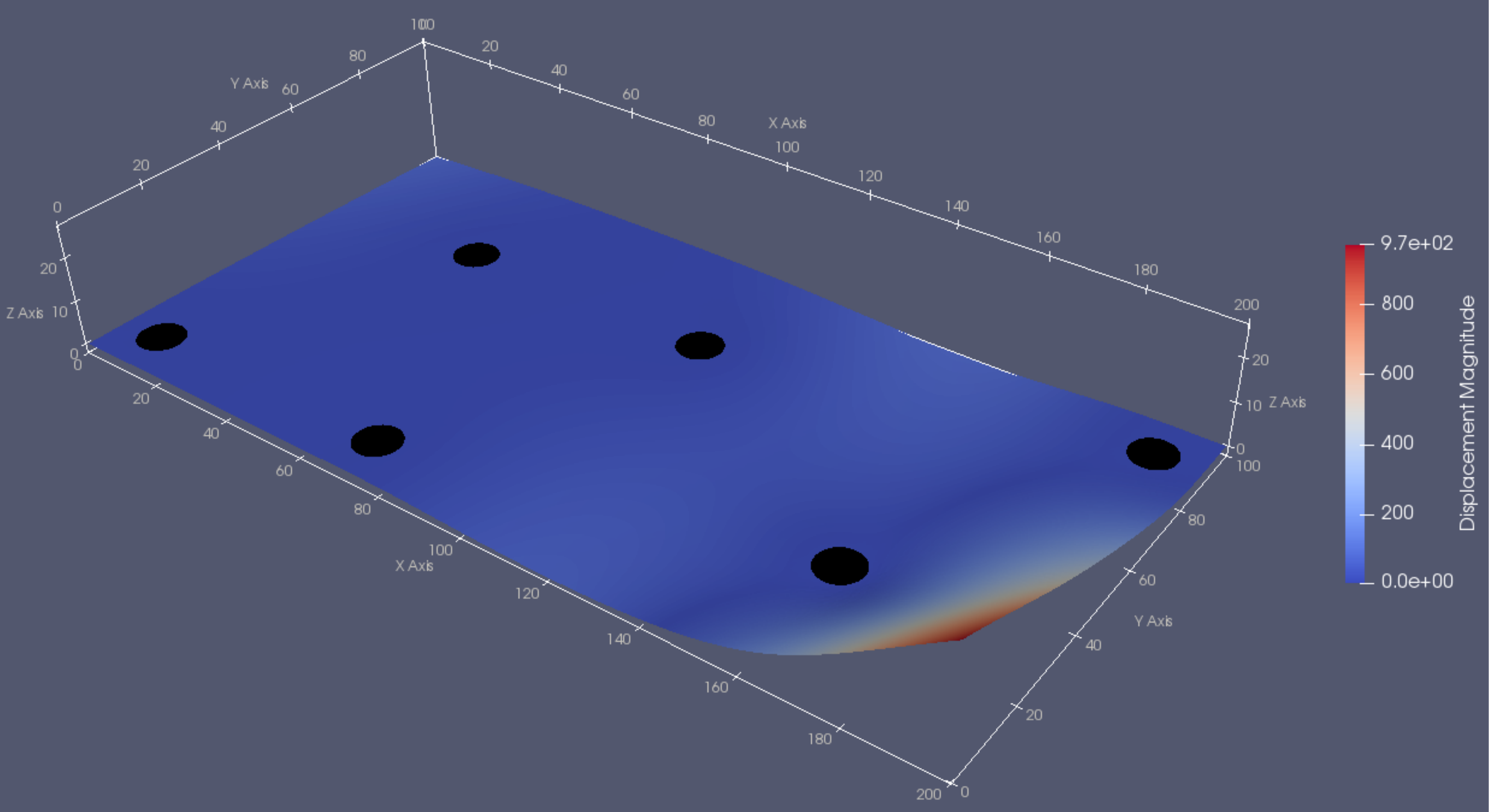}
		\subcaption{The QUBO in Eq. \eqref{eq:qubo_ecm} with $\overline{N}=6$}
		\label{fig:17var_ecm_opt_mode}
	\end{minipage}
	\caption{The frequency analysis results with the optimal mounting hole positions (black circles) shown in Figs. \ref{fig:17var_wsm_opt} and \ref{fig:17var_ecm_opt}. The color contours illustrate the displacement magnitude. }\label{fig:17var_mode_shape}
\end{figure}

For the frequency analysis, we employed the open-source finite-element analysis software, CalculiX \cite{dhondt2004finite}.
The finite element model of the PCB, which we used in Secs. \ref{subsec:wsm} and \ref{subsec:ecm}, was discretized by a quadratic triangular shell element, and its total number of nodes and elements were 19,113 and 9,406, respectively.
We only considered two materials corresponding to the substrate and the mass component in the PCB model.
Domains, $D_{i}^\text{fix}, D_{l}^\text{mass}$, were placed on the discretized PCB model in advance to assign the mounting holes and the additional masses.
In the optimization, the boundary conditions for fixing the displacement and rotation are assigned to $D_{i}^\text{fix}$ where the corresponding design variable $x_i$ becomes 1.
The domains $D_{j}^\text{fix}$, in which $x_j=0$, is regarded as the substrate.
The density in $D_{l}^\text{mass}$ is set to a higher value of the substrate to represent the additional mass, while Young's modulus and Poisson ratio were set to the same value.
Figure \ref{fig:17var_mode_shape} illustrates the frequency analysis result of the optimal solutions, as shown in Figs. \ref{fig:17var_wsm_opt} and \ref{fig:17var_ecm_opt}.

\subsection{QUBO solver}

Here, we used simulated annealing (SA) \cite{kirkpatrick1983optimization} and quantum annealing (QA) \cite{kadowaki1998quantum} as the QUBO solver.
Both SA and QA are heuristic optimization methods, based on the analogy between the search process in the optimization and the physical phenomena.
SA was performed on a classical computer, while QA on a quantum computer that is developed by D-Wave Systems Inc. \cite{johnson2011quantum}.
We performed SA as implemented in dwave-neal \cite{neal}, and QA as in D-Wave Hybrid Solver \cite{HHS}.
In FM-QUBO and BOCS-QUBO, SA was performed 100 times and QA 3000 times in an optimization calculation.
The best solution of the optimized solutions was regarded as the optimal solution and used for constructing the approximation model of FM or BOCS.

\bibliography{bbo_qubo_ref}

\begin{thebibliography}{10}
\urlstyle{rm}
\expandafter\ifx\csname url\endcsname\relax
  \def\url#1{\texttt{#1}}\fi
\expandafter\ifx\csname urlprefix\endcsname\relax\def\urlprefix{URL }\fi
\expandafter\ifx\csname doiprefix\endcsname\relax\def\doiprefix{DOI: }\fi
\providecommand{\bibinfo}[2]{#2}
\providecommand{\eprint}[2][]{\url{#2}}

\bibitem{kadowaki1998quantum}
\bibinfo{author}{Kadowaki, T.} \& \bibinfo{author}{Nishimori, H.}
\newblock \bibinfo{journal}{\bibinfo{title}{{Quantum annealing in the
  transverse Ising model}}}.
\newblock {\emph{\JournalTitle{Physical Review E}}}
  \textbf{\bibinfo{volume}{58}}, \bibinfo{pages}{5355} (\bibinfo{year}{1998}).

\bibitem{das2008colloquium}
\bibinfo{author}{Das, A.} \& \bibinfo{author}{Chakrabarti, B.~K.}
\newblock \bibinfo{journal}{\bibinfo{title}{{Colloquium: Quantum annealing and
  analog quantum computation}}}.
\newblock {\emph{\JournalTitle{Reviews of Modern Physics}}}
  \textbf{\bibinfo{volume}{80}}, \bibinfo{pages}{1061} (\bibinfo{year}{2008}).

\bibitem{ray1989sherrington}
\bibinfo{author}{Ray, P.}, \bibinfo{author}{Chakrabarti, B.~K.} \&
  \bibinfo{author}{Chakrabarti, A.}
\newblock \bibinfo{journal}{\bibinfo{title}{{Sherrington-Kirkpatrick model in a
  transverse field: Absence of replica symmetry breaking due to quantum
  fluctuations}}}.
\newblock {\emph{\JournalTitle{Physical Review B}}}
  \textbf{\bibinfo{volume}{39}}, \bibinfo{pages}{11828} (\bibinfo{year}{1989}).

\bibitem{johnson2011quantum}
\bibinfo{author}{Johnson, M.~W.} \emph{et~al.}
\newblock \bibinfo{journal}{\bibinfo{title}{Quantum annealing with manufactured
  spins}}.
\newblock {\emph{\JournalTitle{Nature}}} \textbf{\bibinfo{volume}{473}},
  \bibinfo{pages}{194--198} (\bibinfo{year}{2011}).

\bibitem{yarkoni2021quantum}
\bibinfo{author}{Yarkoni, S.}, \bibinfo{author}{Raponi, E.},
  \bibinfo{author}{Schmitt, S.} \& \bibinfo{author}{B{\"a}ck, T.}
\newblock \bibinfo{journal}{\bibinfo{title}{Quantum annealing for industry
  applications: Introduction and review}}.
\newblock {\emph{\JournalTitle{arXiv preprint arXiv:2112.07491}}}
  (\bibinfo{year}{2021}).

\bibitem{tsukamoto2017accelerator}
\bibinfo{author}{Tsukamoto, S.}, \bibinfo{author}{Takatsu, M.},
  \bibinfo{author}{Matsubara, S.} \& \bibinfo{author}{Tamura, H.}
\newblock \bibinfo{journal}{\bibinfo{title}{An accelerator architecture for
  combinatorial optimization problems}}.
\newblock {\emph{\JournalTitle{{FUJITSU Science and Technology Journal}}}}
  \textbf{\bibinfo{volume}{53}}, \bibinfo{pages}{8--13} (\bibinfo{year}{2017}).

\bibitem{goto2019combinatorial}
\bibinfo{author}{Goto, H.}, \bibinfo{author}{Tatsumura, K.} \&
  \bibinfo{author}{Dixon, A.~R.}
\newblock \bibinfo{journal}{\bibinfo{title}{Combinatorial optimization by
  simulating adiabatic bifurcations in nonlinear {Hamiltonian} systems}}.
\newblock {\emph{\JournalTitle{Science Advances}}}
  \textbf{\bibinfo{volume}{5}}, \bibinfo{pages}{eaav2372}
  (\bibinfo{year}{2019}).

\bibitem{yoshimura2013spatial}
\bibinfo{author}{Yoshimura, C.}, \bibinfo{author}{Yamaoka, M.},
  \bibinfo{author}{Aoki, H.} \& \bibinfo{author}{Mizuno, H.}
\newblock \bibinfo{title}{Spatial computing architecture using randomness of
  memory cell stability under voltage control}.
\newblock In \emph{\bibinfo{booktitle}{2013 European Conference on Circuit
  Theory and Design (ECCTD)}}, \bibinfo{pages}{1--4}
  (\bibinfo{organization}{IEEE}, \bibinfo{year}{2013}).

\bibitem{inagaki2016coherent}
\bibinfo{author}{Inagaki, T.} \emph{et~al.}
\newblock \bibinfo{journal}{\bibinfo{title}{A coherent ising machine for
  2000-node optimization problems}}.
\newblock {\emph{\JournalTitle{Science}}} \textbf{\bibinfo{volume}{354}},
  \bibinfo{pages}{603--606} (\bibinfo{year}{2016}).

\bibitem{amplify}
\bibinfo{author}{Amplify, F.}
\newblock \bibinfo{title}{Fixstars {Amplify} software}.
\newblock \bibinfo{note}{\url{https://amplify.fixstars.com/en/docs/}.}

\bibitem{irie2021}
\bibinfo{author}{Irie, H.}, \bibinfo{author}{Liang, H.}, \bibinfo{author}{Doi,
  T.}, \bibinfo{author}{Gongyo, S.} \& \bibinfo{author}{Hatsuda, T.}
\newblock \bibinfo{journal}{\bibinfo{title}{Hybrid quantum annealing via
  molecular dynamics}}.
\newblock {\emph{\JournalTitle{Scientific Reports}}}
  \textbf{\bibinfo{volume}{11}}, \bibinfo{pages}{1--9} (\bibinfo{year}{2021}).

\bibitem{ohzeki2019control}
\bibinfo{author}{Ohzeki, M.}, \bibinfo{author}{Miki, A.},
  \bibinfo{author}{Miyama, M.~J.} \& \bibinfo{author}{Terabe, M.}
\newblock \bibinfo{journal}{\bibinfo{title}{Control of automated guided
  vehicles without collision by quantum annealer and digital devices}}.
\newblock {\emph{\JournalTitle{Frontiers in Computer Science}}}
  \textbf{\bibinfo{volume}{1}}, \bibinfo{pages}{9} (\bibinfo{year}{2019}).

\bibitem{lucas2014ising}
\bibinfo{author}{Lucas, A.}
\newblock \bibinfo{journal}{\bibinfo{title}{Ising formulations of many {NP}
  problems}}.
\newblock {\emph{\JournalTitle{Frontiers in Physics}}}
  \textbf{\bibinfo{volume}{2}}, \bibinfo{pages}{5} (\bibinfo{year}{2014}).

\bibitem{chancellor2019domain}
\bibinfo{author}{Chancellor, N.}
\newblock \bibinfo{journal}{\bibinfo{title}{Domain wall encoding of discrete
  variables for quantum annealing and {QAOA}}}.
\newblock {\emph{\JournalTitle{Quantum Science and Technology}}}
  \textbf{\bibinfo{volume}{4}}, \bibinfo{pages}{045004} (\bibinfo{year}{2019}).

\bibitem{tamura2021performance}
\bibinfo{author}{Tamura, K.}, \bibinfo{author}{Shirai, T.},
  \bibinfo{author}{Katsura, H.}, \bibinfo{author}{Tanaka, S.} \&
  \bibinfo{author}{Togawa, N.}
\newblock \bibinfo{journal}{\bibinfo{title}{Performance comparison of typical
  binary-integer encodings in an {Ising} machine}}.
\newblock {\emph{\JournalTitle{IEEE Access}}} \textbf{\bibinfo{volume}{9}},
  \bibinfo{pages}{81032--81039} (\bibinfo{year}{2021}).

\bibitem{dattani2019quadratization}
\bibinfo{author}{Dattani, N.}
\newblock \bibinfo{journal}{\bibinfo{title}{Quadratization in discrete
  optimization and quantum mechanics}}.
\newblock {\emph{\JournalTitle{arXiv preprint arXiv:1901.04405}}}
  (\bibinfo{year}{2019}).

\bibitem{neukart2017traffic}
\bibinfo{author}{Neukart, F.} \emph{et~al.}
\newblock \bibinfo{journal}{\bibinfo{title}{Traffic flow optimization using a
  quantum annealer}}.
\newblock {\emph{\JournalTitle{Frontiers in ICT}}}
  \textbf{\bibinfo{volume}{4}}, \bibinfo{pages}{29} (\bibinfo{year}{2017}).

\bibitem{inoue2021traffic}
\bibinfo{author}{Inoue, D.}, \bibinfo{author}{Okada, A.},
  \bibinfo{author}{Matsumori, T.}, \bibinfo{author}{Aihara, K.} \&
  \bibinfo{author}{Yoshida, H.}
\newblock \bibinfo{journal}{\bibinfo{title}{Traffic signal optimization on a
  square lattice with quantum annealing}}.
\newblock {\emph{\JournalTitle{Scientific reports}}}
  \textbf{\bibinfo{volume}{11}}, \bibinfo{pages}{1--12} (\bibinfo{year}{2021}).

\bibitem{rosenberg2016solving}
\bibinfo{author}{Rosenberg, G.} \emph{et~al.}
\newblock \bibinfo{journal}{\bibinfo{title}{Solving the optimal trading
  trajectory problem using a quantum annealer}}.
\newblock {\emph{\JournalTitle{IEEE Journal of Selected Topics in Signal
  Processing}}} \textbf{\bibinfo{volume}{10}}, \bibinfo{pages}{1053--1060}
  (\bibinfo{year}{2016}).

\bibitem{streif2019solving}
\bibinfo{author}{Streif, M.}, \bibinfo{author}{Neukart, F.} \&
  \bibinfo{author}{Leib, M.}
\newblock \bibinfo{title}{Solving quantum chemistry problems with a {D-Wave}
  quantum annealer}.
\newblock In \emph{\bibinfo{booktitle}{International Workshop on Quantum
  Technology and Optimization Problems}}, \bibinfo{pages}{111--122}
  (\bibinfo{organization}{Springer}, \bibinfo{year}{2019}).

\bibitem{amin2018quantum}
\bibinfo{author}{Amin, M.~H.}, \bibinfo{author}{Andriyash, E.},
  \bibinfo{author}{Rolfe, J.}, \bibinfo{author}{Kulchytskyy, B.} \&
  \bibinfo{author}{Melko, R.}
\newblock \bibinfo{journal}{\bibinfo{title}{Quantum boltzmann machine}}.
\newblock {\emph{\JournalTitle{Physical Review X}}}
  \textbf{\bibinfo{volume}{8}}, \bibinfo{pages}{021050} (\bibinfo{year}{2018}).

\bibitem{bendsoe1988generating}
\bibinfo{author}{Bends{\o}e, M.~P.} \& \bibinfo{author}{Kikuchi, N.}
\newblock \bibinfo{journal}{\bibinfo{title}{Generating optimal topologies in
  structural design using a homogenization method}}.
\newblock {\emph{\JournalTitle{Computer Methods in Applied Mechanics and
  Engineering}}} \textbf{\bibinfo{volume}{71}}, \bibinfo{pages}{197--224}
  (\bibinfo{year}{1988}).

\bibitem{deaton2014survey}
\bibinfo{author}{Deaton, J.~D.} \& \bibinfo{author}{Grandhi, R.~V.}
\newblock \bibinfo{journal}{\bibinfo{title}{A survey of structural and
  multidisciplinary continuum topology optimization: post 2000}}.
\newblock {\emph{\JournalTitle{Structural and Multidisciplinary Optimization}}}
  \textbf{\bibinfo{volume}{49}}, \bibinfo{pages}{1--38} (\bibinfo{year}{2014}).

\bibitem{kitai2020designing}
\bibinfo{author}{Kitai, K.} \emph{et~al.}
\newblock \bibinfo{journal}{\bibinfo{title}{Designing metamaterials with
  quantum annealing and factorization machines}}.
\newblock {\emph{\JournalTitle{Physical Review Research}}}
  \textbf{\bibinfo{volume}{2}}, \bibinfo{pages}{013319} (\bibinfo{year}{2020}).

\bibitem{wilson2021machine}
\bibinfo{author}{Wilson, B.~A.} \emph{et~al.}
\newblock \bibinfo{journal}{\bibinfo{title}{Machine learning framework for
  quantum sampling of highly-constrained, continuous optimization problems}}.
\newblock {\emph{\JournalTitle{arXiv preprint arXiv:2105.02396}}}
  (\bibinfo{year}{2021}).

\bibitem{izawa2021continuous}
\bibinfo{author}{Izawa, S.}, \bibinfo{author}{Kitai, K.},
  \bibinfo{author}{Tanaka, S.}, \bibinfo{author}{Tamura, R.} \&
  \bibinfo{author}{Tsuda, K.}
\newblock \bibinfo{journal}{\bibinfo{title}{Continuous black-box optimization
  with quantum annealing and random subspace coding}}.
\newblock {\emph{\JournalTitle{arXiv preprint arXiv:2104.14778}}}
  (\bibinfo{year}{2021}).

\bibitem{koshikawa2021benchmark}
\bibinfo{author}{Koshikawa, A.~S.}, \bibinfo{author}{Ohzeki, M.},
  \bibinfo{author}{Kadowaki, T.} \& \bibinfo{author}{Tanaka, K.}
\newblock \bibinfo{journal}{\bibinfo{title}{{Benchmark test of Black-box
  optimization using D-Wave quantum annealer}}}.
\newblock {\emph{\JournalTitle{Journal of the Physical Society of Japan}}}
  \textbf{\bibinfo{volume}{90}}, \bibinfo{pages}{064001}
  (\bibinfo{year}{2021}).

\bibitem{rendle2010factorization}
\bibinfo{author}{Rendle, S.}
\newblock \bibinfo{title}{Factorization machines}.
\newblock In \emph{\bibinfo{booktitle}{IEEE International Conference on Data
  Mining}}, \bibinfo{pages}{995--1000} (\bibinfo{organization}{IEEE},
  \bibinfo{year}{2010}).

\bibitem{baptista2018bayesian}
\bibinfo{author}{Baptista, R.} \& \bibinfo{author}{Poloczek, M.}
\newblock \bibinfo{title}{Bayesian optimization of combinatorial structures}.
\newblock In \emph{\bibinfo{booktitle}{International Conference on Machine
  Learning}}, \bibinfo{pages}{462--471} (\bibinfo{year}{2018}).

\bibitem{hatakeyama2021tackling}
\bibinfo{author}{Hatakeyama-Sato, K.}, \bibinfo{author}{Kashikawa, T.},
  \bibinfo{author}{Kimura, K.} \& \bibinfo{author}{Oyaizu, K.}
\newblock \bibinfo{journal}{\bibinfo{title}{Tackling the challenge of a huge
  materials science search space with quantum-inspired annealing}}.
\newblock {\emph{\JournalTitle{Advanced Intelligent Systems}}}
  \textbf{\bibinfo{volume}{3}}, \bibinfo{pages}{2000209}
  (\bibinfo{year}{2021}).

\bibitem{koshikawa2021combinatorial}
\bibinfo{author}{Koshikawa, A.~S.} \emph{et~al.}
\newblock \bibinfo{journal}{\bibinfo{title}{Combinatorial black-box
  optimization for vehicle design problem}}.
\newblock {\emph{\JournalTitle{arXiv preprint arXiv:2110.00226}}}
  (\bibinfo{year}{2021}).

\bibitem{deb2001}
\bibinfo{author}{Deb, K.}
\newblock \emph{\bibinfo{title}{Multi-objective optimization using evolutionary
  algorithms}} (\bibinfo{publisher}{{John Wiley \& Sons}},
  \bibinfo{year}{2001}).

\bibitem{rendle2012factorization}
\bibinfo{author}{Rendle, S.}
\newblock \bibinfo{journal}{\bibinfo{title}{Factorization machines with
  libfm}}.
\newblock {\emph{\JournalTitle{ACM Transactions on Intelligent Systems and
  Technology}}} \textbf{\bibinfo{volume}{3}}, \bibinfo{pages}{1--22}
  (\bibinfo{year}{2012}).

\bibitem{morita2006convergence}
\bibinfo{author}{Morita, S.} \& \bibinfo{author}{Nishimori, H.}
\newblock \bibinfo{journal}{\bibinfo{title}{Convergence theorems for quantum
  annealing}}.
\newblock {\emph{\JournalTitle{Journal of Physics A: Mathematical and
  General}}} \textbf{\bibinfo{volume}{39}}, \bibinfo{pages}{13903}
  (\bibinfo{year}{2006}).

\bibitem{dhondt2004finite}
\bibinfo{author}{Dhondt, G.}
\newblock \emph{\bibinfo{title}{The finite element method for three-dimensional
  thermomechanical applications}} (\bibinfo{publisher}{John Wiley \& Sons},
  \bibinfo{year}{2004}).

\bibitem{kirkpatrick1983optimization}
\bibinfo{author}{Kirkpatrick, S.}, \bibinfo{author}{Gelatt, C.~D.} \&
  \bibinfo{author}{Vecchi, M.~P.}
\newblock \bibinfo{journal}{\bibinfo{title}{Optimization by simulated
  annealing}}.
\newblock {\emph{\JournalTitle{Science}}} \textbf{\bibinfo{volume}{220}},
  \bibinfo{pages}{671--680}, \doiprefix\url{10.1126/science.220.4598.671}
  (\bibinfo{year}{1983}).

\bibitem{neal}
\bibinfo{author}{{D-Wave Systems Inc.}}
\newblock \bibinfo{title}{dwave-neal}.
\newblock \bibinfo{note}{\url{https://github.com/dwavesystems/dwave-neal}.}

\bibitem{HHS}
\bibinfo{author}{{D-Wave Systems Inc.}}
\newblock \bibinfo{title}{{D-Wave} hybrid solver service: An overview}.
\newblock
  \bibinfo{note}{\url{https://www.dwavesys.com/sites/default/files/14-1039A-A_D-Wave_Hybrid_Solver_Service_An_Overview.pdf}.}

\end{thebibliography}

\section*{Author contributions}

T.M. M.T. and T.K. conceived and developed the concept. T.M. carried out all the experiments, analyzed the results, and wrote the
manuscript. All authors reviewed the manuscript.

\end{document}